\documentclass[aps,prb,preprint,superscriptaddress,amsmath,amssymb]{revtex4-2}

\usepackage{graphicx}
\usepackage{lipsum}

\usepackage{hyperref}  
\usepackage[all]{hypcap} 

\usepackage{color}

\newcommand{\chem}[1]{\ensuremath{\mathrm{#1}}} 
\newcommand{\co}{Co_3O_2BO_3}

\usepackage{float}

\usepackage[normalem]{ulem}

\begin{document}

\title{Spin-State Ordering and Intermediate States in the Mixed-Valence Cobalt Oxyborate Co$_3$O$_2$BO$_3$ with Spin Crossover}

\author{E. Granado}
\email[]{granado@unicamp.br}
\affiliation{Gleb Wataghin Institute of Physics, University of Campinas (UNICAMP), Campinas, S\~ao Paulo, 13083-859, Brazil}

\author{C. W. Galdino}
\altaffiliation[Present address: ] {Swiss Light Source, Paul Scherrer Institute, 5232 Villigen-PSI, Switzerland}
\affiliation{Gleb Wataghin Institute of Physics, University of Campinas (UNICAMP), Campinas, S\~ao Paulo, 13083-859, Brazil}

\author{B. D. Moreno}
\affiliation{Canadian Light Source Inc., 44 Innovation Boulevard, Saskatoon, SK S7N 2V3, Canada}

\author{G. King}
\affiliation{Canadian Light Source Inc., 44 Innovation Boulevard, Saskatoon, SK S7N 2V3, Canada}

\author{D. C. Freitas}
\affiliation{Instituto de F\'isica, Universidade Federal Fluminense, Campus da Praia Vermelha, Niter\'oi, RJ, 24210-346, Brazil}

\date{\today}

\begin{abstract}

Spin-state ordering - a periodic pattern of ions with different spin-state configurations along a crystal lattice - is a rare phenomenon, and its possible interrelation with other electronic degrees of freedom remains little explored. Here we perform a structural investigation of the mixed-valence Co homometallic ludwigite Co$_2^{2+}$Co$^{3+}$O$_2$BO$_3$. A superstructure consistent with a long-range Co$^{3+}$ spin-state ordering is observed between $T_{4}=580$ K and $T_{3}=510$ K. Intermediate states with mesoscopic correlations are detected below $T_{3}$ down to $T_{1}=480$ K with a change of dimensionality at $T_{2}=495$ K. The spin-state correlations are connected to the charge sector as revealed by the abrupt changes in the electrical resistance at $T_1$ and $T_2$. The evolution of the structural parameters below $T_{1}$ indicate that the spin crossover is ignited by a moderate degree of thermally-induced Co$^{2+}$ and Co$^{3+}$ charge disorder. Charge and spin-state degrees of freedom can be interrelated in mixed-valence spin-crossover materials, leading to sharp transitions involving intermediate spin-state and charge-correlated states at the mesoscale. 

\end{abstract}

\maketitle

\newpage

\section{Introduction}

Transition-metal compounds often display intertwined charge, orbital and spin degrees of freedom \cite{Tranquada1995,Khomskii1997,Salamon2001,Zhang2020,Teng2022}. In addition, some ions such as Co$^{3+}$ may have competing atomic spin-orbital configurations, namely low-spin (LS), high-spin (HS), and arguably intermediate-spin states \cite{korotin1996intermediate,Lamonova2011}. In spin crossover materials, different spin states may coexist over broad temperature intervals \cite{haverkort2006spin,cheng2012spin,kazak2021spin}, and in some of them the coexisting spin states order in the crystal lattice forming a superstructure \cite{Doumerc1999,Doumerc2001,khomskii2004superstructures,Collet2012,Hu_2012,Ikeda2016,Murnaghan2014}. If the metallic ions are also mixed-valent, spin-state and charge orders may in principle coexist and either compete or support each order, potentially leading to a peculiar type of spin-charge coupling that remains to be experimentally confirmed. Arguably, high-quality crystalline samples with relatively complex structures allowing for multiple transition-metal sites are needed to host such rich physics, and detailed structural investigations with the aid of supporting physical property experiments are necessary to identify it.

The $M_2^{2+}M'^{3+}$O$_2$BO$_3$ oxyborates with the ludwigite structure present metallic ions $M^{2+}$ and $M'^{3+}$ in different oxidation states and at four crystallographic sites $M(1)-M(4)$, leading to a large variety of possibilities for chemical, charge, and magnetic orders \cite{takeuchi1950crystal,wiedenmann1978magnetic,norrestam1994structural,fernandes1998magnetic,irwin1999crystal,guimaraes1999cation,continentino1999magnetic,fernandes2000specific,mir2001structural,continentino2001magnetic,whangbo2002theoretical,sanchez2004magnetism,borates2005frontiers,fernandes2005transport,mir2006x,ivanova2007magnetic,freitas2008structure,bordet2009magnetic,freitas2009partial,kazak2009conductivity,kazak2009low,freitas2010structural,kazak2011superexchange,bartolome2011uniaxial,ivanova2011crystal,knyazev2012crystal,ivanova2012effect,ivanova2012spin,ivanova2013structure,medrano2015nonmagnetic,dos2016current,freitas2016magnetism,sofronova2017ludwigites,medrano2017,kumar2017reentrant,dos2017non,galdino2019magnetic,knyazev2019effect,mariano2020,heringer2020spin,kazak2021spin,medrano2021magnetic,galdino2021,kazak2022}. The crystal structure is composed of an arrangement of edge- and corner-shared $M$O$_6$ and $M'$O$_6$ octahedra that are stacked along {\bf c} \cite{takeuchi1950crystal,irwin1999crystal,sofronova2017ludwigites}. It is often convenient to describe this structure in terms of subunits termed three-legged ladders (3LL), composed of $M(3)-M(1)-M(3)$ and $M(4)-M(2)-M(4)$ metallic ions (313 and 424, respectively) \cite{mir2001structural,sofronova2017ludwigites}. For homometallic ludwigites ($M=M'$), different valence states of $M$ coexist, and charge order may become an important ingredient of the physics of these materials \cite{mir2001structural,bordet2009magnetic,freitas2016magnetism}. For instance, Fe$_3$O$_2$BO$_3$ shows a structural transition at $T_c=283$ K, where the low-temperature and lower-symmetry phase ($Pbnm$ space group) shows a splitting of the Fe(4) site into non-equivalent Fe($4a$) and Fe($4b$). The Fe($4a$)-Fe(2) distance is substantially smaller than Fe($4b$)-Fe(2), and this transition was ascribed to a dimerized charge-ordered state below $T_c$ \cite{mir2001structural,mir2006x,bordet2009magnetic}.

Co$_3$O$_2$BO$_3$, the other known homometallic ludwigite, do not display any symmetry reduction with respect to the regular ludwigite structure at low temperatures \cite{freitas2008structure,freitas2016magnetism,galdino2019magnetic,galdino2021,kazak2021spin}. This is because its ground-state charge configuration is a trivial columnar charge-ordered state, with the Co$^{3+}$ ions fully occupying a single crystallographic Co(4) site without the need of further symmetry reduction with respect to the parent ludwigite structure. Another important ingredient of this material is the spin state of Co$^{3+}$. The magnetic structure below $T_N=43$ K corroborates a LS Co$^{3+}$ ground state configuration \cite{freitas2016magnetism}. On the other hand, the relatively large value of the high-temperature paramagnetic moment of this material indicates that at least part of the Co$^{3+}$ ions is converted into a higher-spin state upon warming \cite{freitas2008structure,galdino2019magnetic,kazak2021spin}. Indeed, former studies of the physical properties and crystal structure of Co$_3$O$_2$BO$_3$ indicate non-trivial phenomena occurring above room temperature \cite{galdino2019magnetic,kazak2021spin,galdino2021,kazak2022}. Substantial anomalies in the lattice parameters and Co-O bond distances are observed between $\sim 370$ and $\sim 550$ K \cite{galdino2019magnetic,kazak2021spin,kazak2022}. Also, bulk phase transitions at $T_{1} \sim 480$ K and $T_{2}=495$ K were evidenced by electrical conductivity and differential scanning calorimetry measurements \cite{galdino2019magnetic,galdino2021}. In this material, both charge and spin-state degrees of freedom are active, and a combination of them is likely needed to understand such a complex physical behavior. 

In this work, synchrotron X-ray diffraction and electrical resistance measurements are performed in Co$_3$O$_2$BO$_3$ using single crystalline and powdered samples of Co$_3$O$_2$BO$_3$. Our results demonstrate a sequence of phase transitions between $T_{1} = 480$ and $T_{4}=580$ K involving a long-range spin-state-ordered state as well as intermediate states with anisotropic correlations in the mesoscale. The degree of charge ordering and the proportion of HS Co$^{3+}$ state is highly temperature dependent above $\sim 370$ K, provoking an electronic instability in the system that ultimately leads to the observed high-temperature phase transitions. 

\section{Experimental details}

Needle-shaped single crystals of Co$_3$O$_2$BO$_3$ employed in X-ray diffraction and electrical measurements, with sizes of $\sim 50 \times 50 \times 200$ $\mu$m and $\sim 50 \times 50 \times 670$ $\mu$m, respectively, were synthesized as described in Ref. \onlinecite{freitas2008structure} and taken from the same batch as those used in previous works \cite{galdino2019magnetic,galdino2021}. The sample employed in the X-ray powder diffraction experiment was obtained by grinding a large number of small crystals. As the crystals were not individually selected before grinding, crystals of Co$_3$O$_4$ that are byproducts of the single crystal synthesis were not removed, leading to an impurity phase of Co$_3$O$_4$ ($1.7$ \% weight fraction) in the so-obtained powdered sample.

The X-ray powder diffraction experiment was performed at the BXDS-WHE beamline of the Brockhouse diffraction sector of the Canadian Light Source with incoming photons of $E=54.74$ keV ($\lambda=0.2265$ \AA). The beamsize at the sample position was $\sim 50$ (V) $\times 200$ (H) $\mu$m$^2$. The sample was loaded into a 0.64~mm-diameter spinning kapton capillary and the temperature was changed between $130$ and $475$~K by means of a N$_2$ cryostream system. The spinning capillary was crucial to minimize non-statistical fluctuations in the Bragg intensities due to poor powder statistics derived from the large crystallite sizes (tens of microns). As the nominal temperature of the nitrogen flow is not necessarily equal to the sample temperature inside the capillary, a temperature calibration was performed using the thermal expansion of the lattice parameters of our sample in comparison to that obtained in a previous work with neutron diffraction data using a furnace under vacuum \cite{galdino2021}. The highest attainable temperature of our powder diffraction experiment, $T=475$ K, correspond to a $T=500$ K nominal temperature of the nitrogen flow. A 2D image plate detector with CsI absorber was employed, with pixel size of $150 \times 150$ $\mu$m$^2$ and active area of $0.432 \times 0.432$ m$^2$ ($2880 \times 2880$ pixels). Sample-to-detector distance was 0.586 m. The Rietveld refinements were performed with the GSAS-II suite \cite{Toby2013}.

The single crystal X-ray diffraction experiment was performed in the BXDS-IUV beamline of the Brockhouse sector of CLS. The photon energy was $E=7.3147$ keV ($\lambda=1.6950$ \AA). The sample, with a natural surface parallel to the (110) plane, was placed over a flat Si crystal resting onto a Linkham cryofurnace (model HFSX350-GI) that was attached to the Eulerian cradle of a Huber 6+2 circle diffractometer. A Pilatus 100K detector was fixed in the $2 \theta$ arm. All Bragg reflections were accessed using the $\omega=0$ mode of the {\it spec}$^{TM}$ control software, in which the incident and diffracted beam directions make the equal angles with respect to the sample (110) surface, avoiding the need for absorption corrections on the relative diffracted intensities for different $hkl$ reflections.

DC-resistivity as a function of temperature was measured in a single-crystal of \chem{\co} along the {\bf c} direction by the four-probe method using a Keithley 2182A Nanovoltmeter and a 6221 Current Source. The crystal was held in the center of an homemade furnace at the Extreme Thermodynamic Conditions Laboratory (LCTE) of the Brazilian synchrotron light source (SIRIUS/LNLS). This allowed us to reach a higher temperature ($T_{max} = 642$ K) compared to our previous report ($T_{max} = 505$ K) \cite{galdino2019magnetic}. The selected current $I=300$ $\mu$A lies within a current interval where the resistivity presents approximately ohmic behavior and the heating/cooling rate was 10 K/min.

All errors indicated in this work are obtained from least-squares fitting procedures and represent one standard deviation.

\section{Results and Analysis}

\subsection{X-ray powder diffraction}

\begin{figure}[!h]
    \centering
    \includegraphics[scale=0.6]{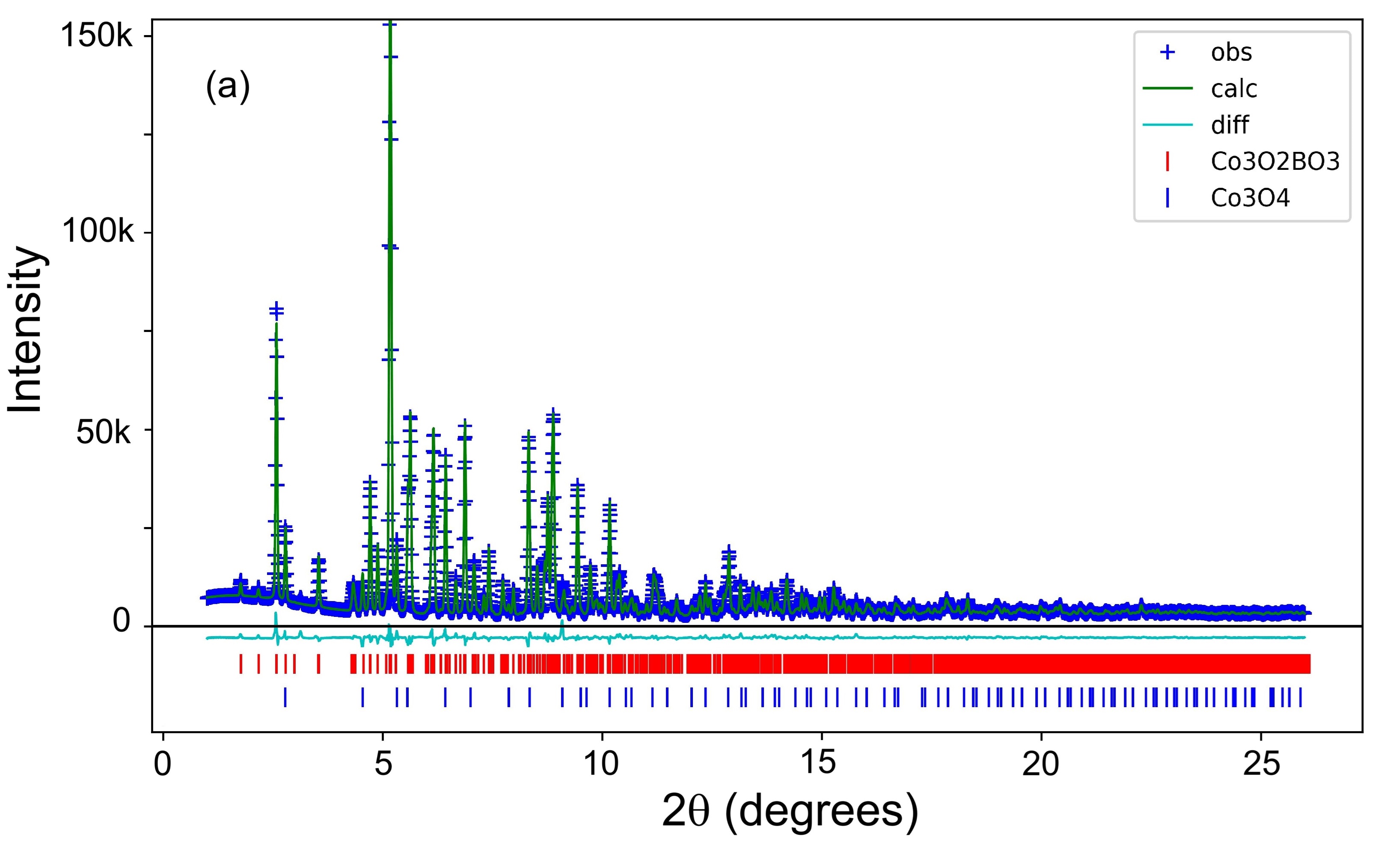}
    \caption{Representative X-ray powder diffraction profile and Rietveld fit at $T=300$ K ($\lambda=0.2265$ \AA, $R_w=4.17 \%$). Cross symbols represent the experimental data and the green solid line is the simulated profile after the refinement. The raw difference data is the solid cyan line. The expected Bragg positions for the main Co$_3$O$_2$BO$_3$ ludwigite and the minor Co$_3$O$_4$ impurity phases (1.7 \% weight fraction) are given in red and blue short vertical bars, respectively.}
    \label{powder}
\end{figure}

\begin{figure}[!h]
    \centering
    \includegraphics[scale=0.35]{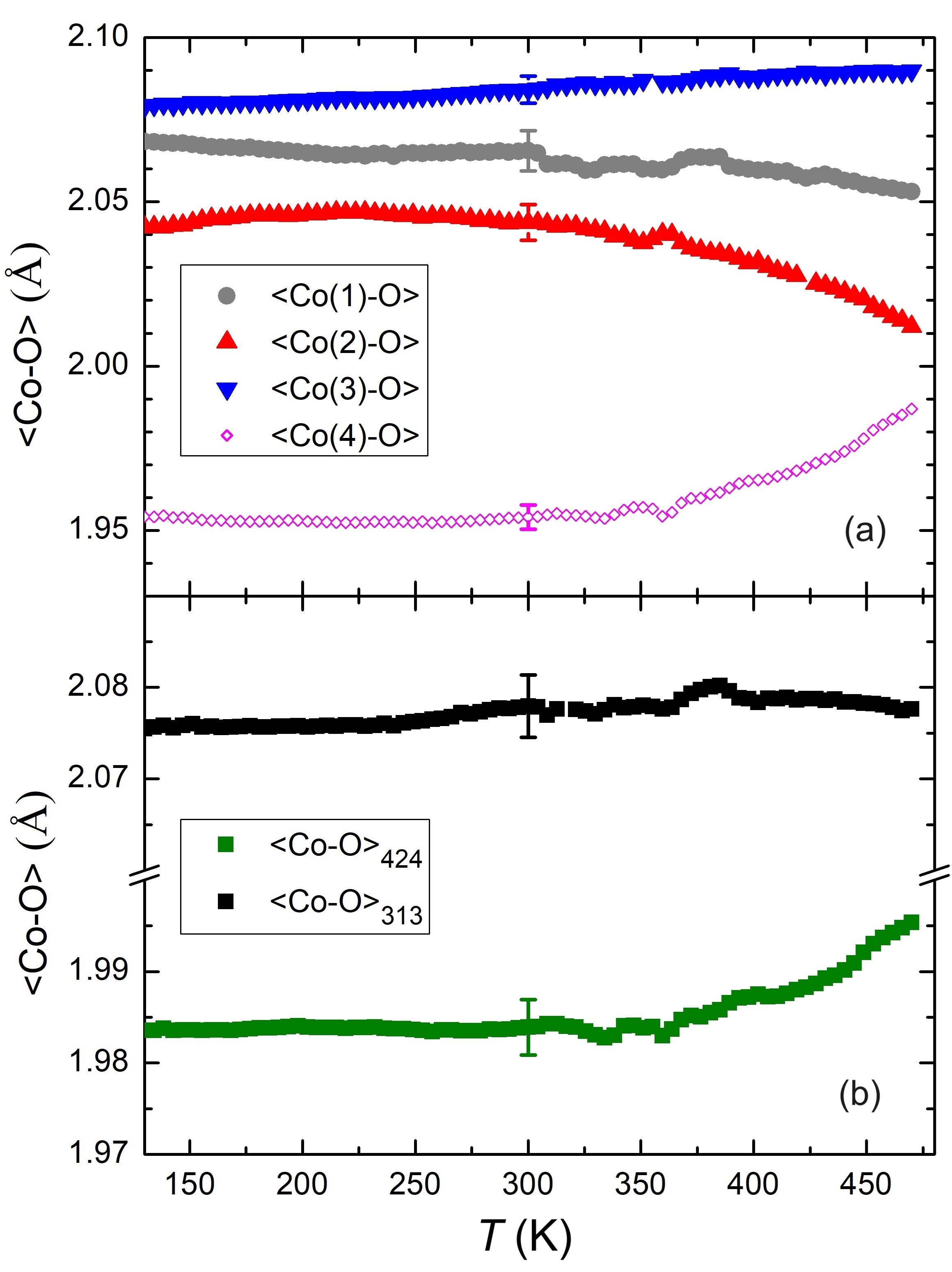}
    \caption{(a) Average $<$Co(n)-O$>$ distances for $n=1-4$ obtained from the refined atomic positions and lattice paramenters. (b) Average $<$Co-O$>$ distances within the 313 and 424 three-legged ladders.}
    \label{distances}
\end{figure}

High-throughput synchrotron X-ray powder diffraction measurements were performed in the temperature interval between $T=125$ and 470 K. Serial structural refinements were carried out using a conventional ludwigite structural model ($Pbam$ space group) \cite{takeuchi1950crystal,irwin1999crystal,freitas2008structure}. The raw data and fits at $T=300$ K are shown in Fig. \ref{powder}. Figure \ref{distances}(a) displays the average $<$Co($n$)-O$>$ distances extracted from the refined structural parameters. The $<$Co(2)-O$>$ is reduced whereas the $<$Co(4)-O$>$ increases upon heating, most notably above $\sim 370$ K, which is attributed to a gradual charge redistribution between these Co sites with respect to the low-temperature ordered configuration of Co(2)$^{2+}$ and Co(4)$^{3+}$ \cite{galdino2021}. Additional insight into the evolution of the electronic states of Co with temperature is gained by plotting the average $<$Co-O$>$ distances within the 313 and 424 3LLs [see Fig. \ref{distances}(b)]. The $<$Co-O$>_{313}$ distance remains nearly constant with $T$, whereas the $<$Co-O$>_{424}$ distance increases upon warming above $\sim 370$ K. We do not attribute this increment of the average Co ionic radius within the 424 3LL to a change of the average oxidation state in this ladder, since this would have to be compensated by changes of opposite sign in the $<$Co-O$>_{313}$ distance, which is not observed. The increment of the $<$Co-O$>_{424}$ distance with $T$ is rather attributed to a modification of the average spin-state of the Co$^{3+}$ ions in the 424 3LL above $\sim 370$ K that accompanies the charge redistribution within this ladder. This is indicative of a charge-driven mechanism for the Co$^{3+}$ spin-state crossover, in which the thermally-induced charge disorder induces a fraction of the Co$^{3+}$ ions, which stays the Co(4) site at low temperatures, to occupy the larger Co(2) site at higher temperatures, thus favoring the conversion of this ion from the LS into a higher-spin state.

\subsection{Single crystal X-ray diffraction}

\begin{figure}
\centering
\includegraphics[scale=0.3]{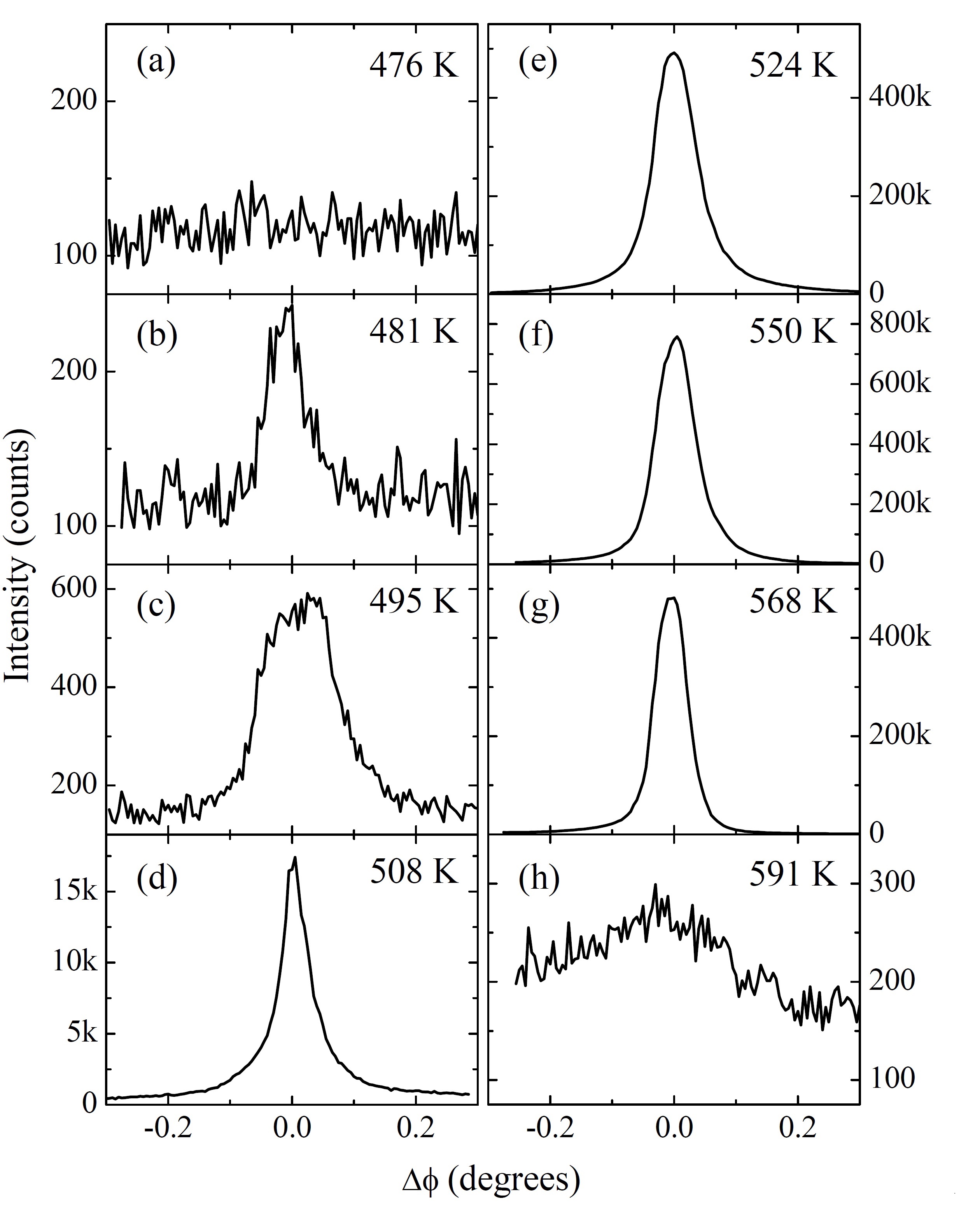}
\caption{(a-h) Azimuthal ($\phi$)-scans of the superstructure 441 reflection ($c$-doubled unit cell) around the [110] axis at selected temperatures.}
\label{phi}
\end{figure}

\begin{figure}
\centering
\includegraphics[scale=0.6]{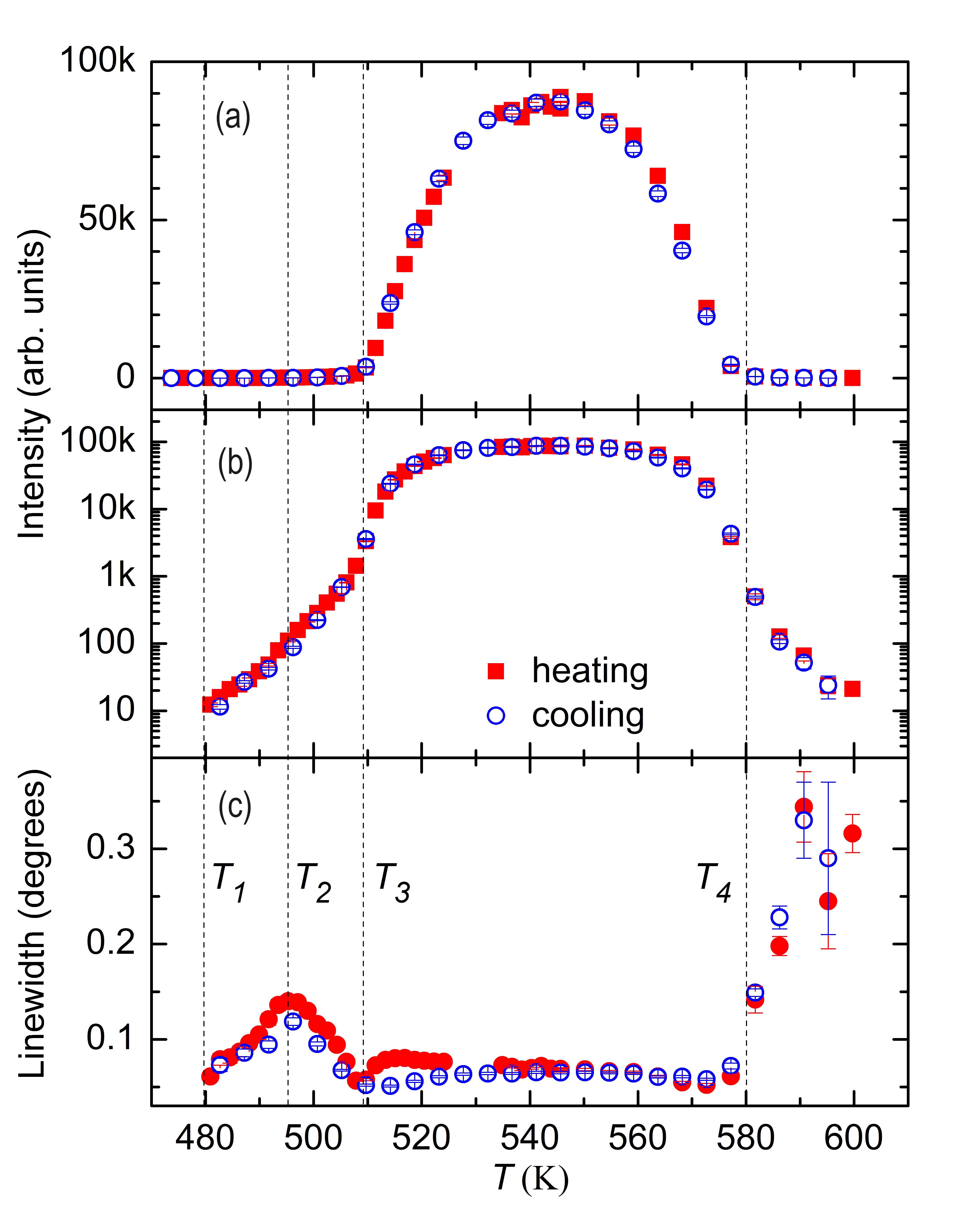}
\caption{Temperature dependence of the integrated intensity of the ($\phi$)-scans in linear (a) and logarithmic (b) scales. The linewidths are given in (c). The identified transition temperatures, $T_1=480$ K, $T_2=495$ K, $T_3=510$ K, and $T_4=580$ K, are marked as vertical dashed lines.}
\label{441}
\end{figure}

Structural transitions are identified by single-crystal X-ray diffraction. Between $T_{3}$ and $T_{4}$, sharp superstructure $hkl$ reflections with half-integer $l$ (considering the standard ludwigite unit cell) are observed. Henceforth, the indexing of these reflections will be given with respect to a $c$-doubled unit cell, so that the superstucture reflections are accordingly indexed with odd-integer $l$. The temperature-dependence of the representative 441 reflection is first investigated by azimuthal scans with the rotation axis $\phi$ being along the [110] crystallographic direction (i.e., perpendicular to the (110) natural sample surface).  Raw scans at selected temperatures are given in Figs. \ref{phi}(a)-\ref{phi}(h). Figures \ref{441}(a) and \ref{441}(b) display the intensity of this peak in linear and logarithmic scales. respectively, whereas the linewidth is shown in Fig. \ref{441}(c). The intensity of this reflection do not evidence any obvious discontinuity with temperature, indicating second-order or very weakly first-order ordering transitions at $T_3=510(1)$ K and $T_4=578(1)$ K. The critical exponent $\beta$ obtained from a fit of the intensities to $I=I_0 |T-T_c|^{2\beta}$ in the ordered phase between $T_3$ and $T_4$ nearby the transitions are $\beta=0.48(10)$ at $T \gtrsim T_3$ and $\beta=0.54(7)$ at $T \lesssim T_4$, which are both consistent with the mean-field value $\beta=1/2$ for a scalar field. Above $T_{4}$ and between $T_{1}$ and $T_{3}$, this peak is still observed in the azimuthal scans, however they are wider, indicating they arise from finite-range correlations. Remarkably, the linewidth has a non-monotonic temperature dependence between $T_{1}$ and $T_{3}$, with maximal broadening at $\sim T_{2}$. No substantial temperature hysteresis is noticed in the area of this peak. Finally, this peak tends to be slightly sharper upon cooling than upon warming.

\begin{figure}
\centering
\includegraphics[scale=0.08]{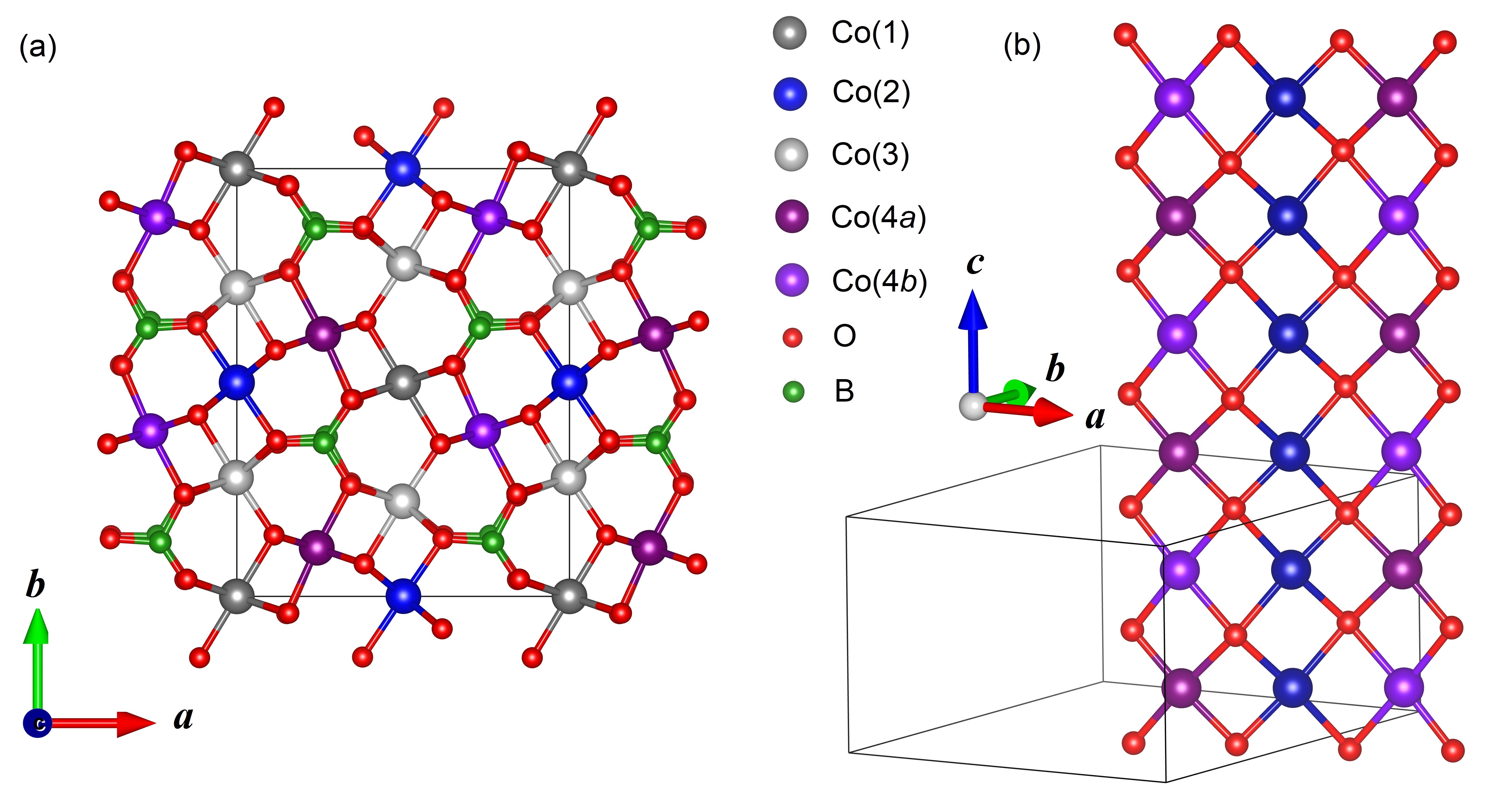}
\caption{(a) {\it ab}-plane projection of the crystal structure of Co$_3$O$_2$BO$_3$ at $T=525$ K ($Pbnm$ space group). (b) Detailed view of the Co($4a$)-Co(2)-Co($4b$) (424) three-legged ladders. The main difference of the superstructure with respect to the regular ludwigite structure ($Pbam$ space group) is the splitting of the Co(4) site of the latter into inequivalent Co($4a$) and Co($4b$) sites that are alternately stacked along {\bf c}, leading to a doubled unit cell [thin black lines in (a) and (b)]. The various oxygen (red) and boron (green) sites are drawn with the same colors for simplicity. This figure was produced using VESTA \cite{momma2011vesta}.}
\label{superstructure}
\end{figure}

The crystal structure of Co$_3$O$_2$BO$_3$ is refined at $T=525$ K, using the integrated intensities of a total of 18 $hkl$ superstructure reflections with odd $l$. The complete list of probed superstructure reflections and their relative intensities are given in Table \ref{hkl}. We also employ the large set of 1480 $hkl$ reflections with even $l$ allowed for the conventional ludwigite structure that was previously obtained at this temperature with conventional X-rays \cite{galdino2021}. The refinement was carried out under the $Pbnm$ space group, which is the same of the related Fe$_3$O$_2$BO$_3$ ludwigite below $T_c=283$ K \cite{mir2001structural}. The refined set of atomic parameters is given in Table \ref{refined}, whereas the relevant Co-O and Co-Co atomic distances extracted from these data are shown in Table \ref{table:distances}. Sketches of the refined crystal structure are drawn in Figs. \ref{superstructure}(a) and \ref{superstructure}(b). The most relevant feature is the splitting of the Co(4) site of the regular ludwigite structure into non-equivalent Co($4a$) and Co($4b$) sites. The interatomic Co($4a$)-Co(2) and Co($4b$)-Co(2) distances are nearly identical [2.7809(1) \AA\ and 2.7814(8) \AA, respectively], indicating that the Co($4a$) and Co($4b$) sites are occupied by ions in the same average oxidation state, also excluding a dimerized state such as proposed for Fe$_3$O$_2$BO$_3$ \cite{mir2001structural}. Remarkably, the $<$Co($4a$)$-$O$>$ and $<$Co($4b$)$-$O$>$ distances are substantially different, [1.966(2) and 2.044(2) \AA, respectively], which is indicative of Co ions with different ionic radii occupying these two sites. Considering that the expected Co-O bond distances derived from the Shannon's radii are 1.945 \AA\ for LS Co$^{3+}$, 2.01 \AA\ for HS Co$^{3+}$, and 2.145 for HS Co$^{2+}$, our results indicate that the Co($4a$) and Co($4b$) sites are mainly occupied by LS and HS Co$^{3+}$ ions, respectively, defining a long-range spin-state-ordered structure between $T_{3}$ and $T_{4}$.

\begin{figure}
\centering
\includegraphics[scale=0.6]{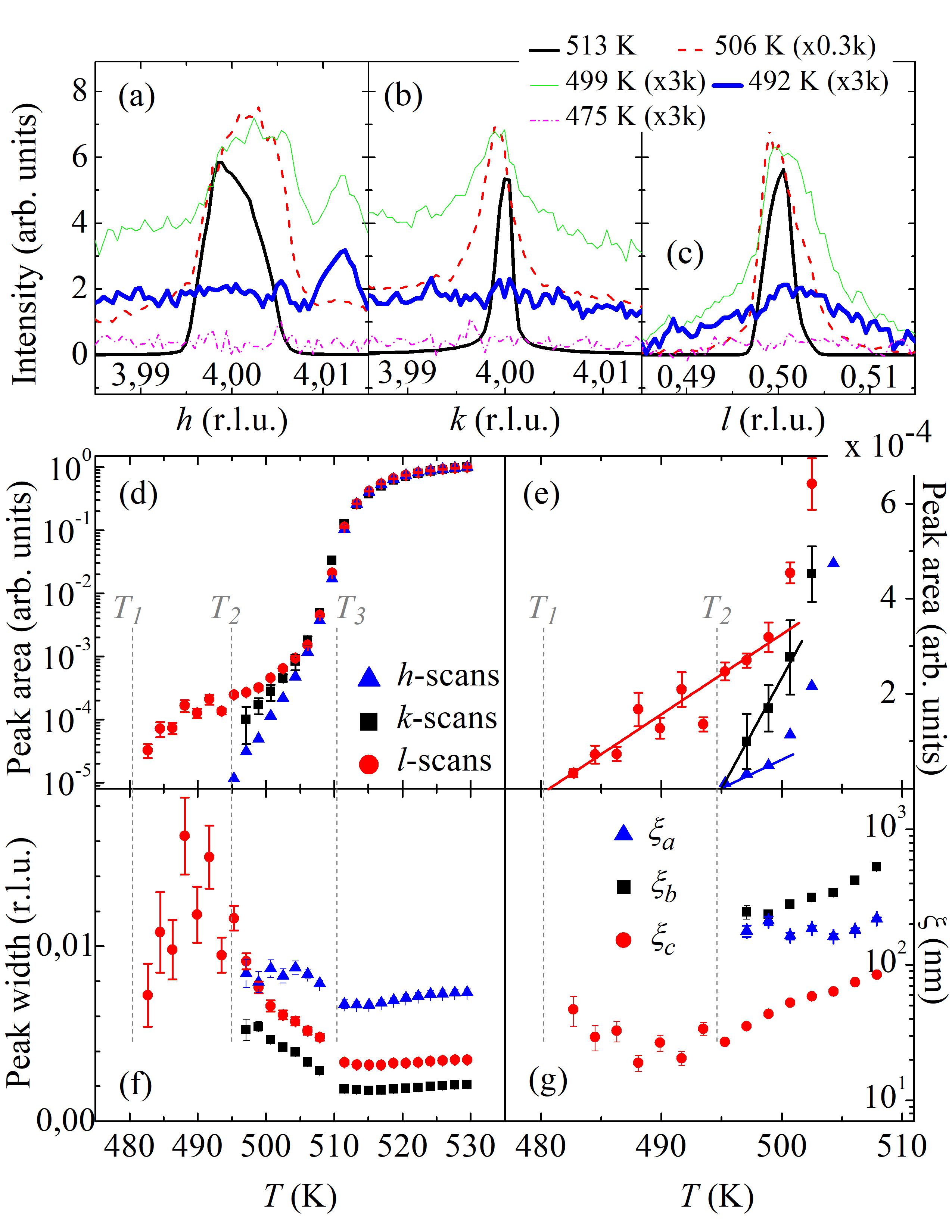}
\caption{Reciprocal-space scans around the 441 superlattice peak along the $h-$ (a), $k-$ (b) and $l-$directions (c) at selected temperatures. (d) Temperature dependence of the 441 peak area in log scale, obtained from the $h-$, $k-$, and $l-$ scans. A detailed view of (d) in linear scale below $\sim 505$ K is given in (e). Solid straight lines are guides to the eyes. (f) Temperature dependence of the 441 peak linewidths of the reciprocal-space scans. The anisotropic correlation lengths $\xi_a$, $\xi_b$, and $\xi_c$, extracted from data of (f) after decovolution of the instrumental widths, are given in (g). The transition temperatures $T_1-T_3$ are marked in (d-g) as vertical dashed lines.}
\label{reciprocal}
\end{figure}

\begin{figure}[!h]
    \centering
    \includegraphics[scale=0.43]{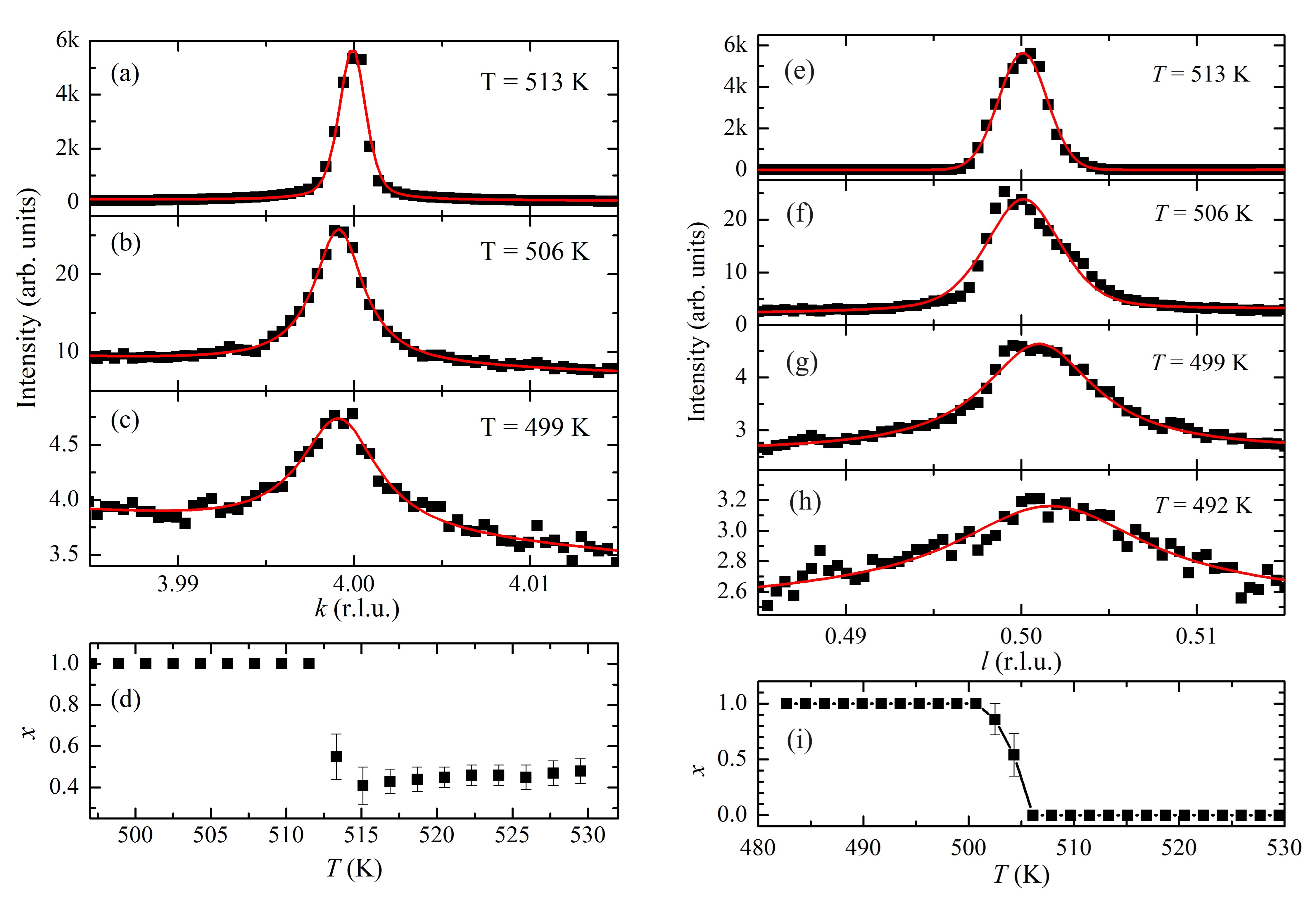}
    \caption{(a-c) Reciprocal-space $k-$ scans around the 441 reflection at $T=513$ K (a), $T=506$ K (b) and $T=499$ K (c) (symbols), and fits to a pseudo-Voigt lineshape, $PV(k-k_0)=xL(k-k_0)+(1-x)G(k-k_0)$ (solid lines), where $L$ and $G$ are Lorentzian and Gaussian functions, respectively, and the Lorentzian weight factor $x$ is restrained to range between 0 and 1. (d) Temperature dependence of $x$ obtained from the fits illustrated in (a-c). (e-h) Reciprocal-space $l-$ scans around the 441 reflection at $T=513$ K (e), $T=506$ K (f), $T=499$ K (g), and $T=492$ K (h) (symbols), and fits to a pseudo-Voigt lineshape (solid lines). (i) Temperature dependence of $x$ obtained from the fits illustrated in (e-h).}
    \label{fits_hk}
\end{figure}

We now focus on the rich behavior of the superstructure peaks between $T_{1}$ and $T_{3}$. One-dimensional reciprocal-space scans across the 441 reflection along the high-symmetry $h$, $k$ and $l$ directions were performed upon cooling below $T=530$ K and are drawn in Figs. \ref{reciprocal}(a)-\ref{reciprocal}(c). 
The observed peaks were fitted with a pseudo-Voigt lineshape, corresponding to a linear combination of Gaussian and Lorentzian functions (see Figs. \ref{fits_hk}(a)-\ref{fits_hk}(i)). Below $T_{3}$, the peak lineshapes for both the $k$- and $l$-scans become fully Lorentzian, indicating that the peak broadening upon cooling below $T_3$ is dominated by the finite size of the spin-state-ordered domains. Thus, the origin of the superstructure peaks below $T_{3}$ is scattering from finite-range spin-state correlations. The temperature dependence of the peak areas are shown in Figs. \ref{reciprocal}(d) and \ref{reciprocal}(e), and the linewidths are given in Fig. \ref{reciprocal}(f). The finite correlation lengths below $T_{3}$ are obtained from the data Fig. \ref{reciprocal}(f) and displayed in Fig. \ref{reciprocal}(g). Between $T_{2}$ and $T_{3}$, the correlation lengths are anisotropic and range between $\sim 20$ and 500 nm. Crucially, between $T_{1}$ and $T_{2}$ the 441 superstructure peak is clearly identified in the $l-$scans and also in the azimuthal scans, but not in the $h-$ and $k-$ scans for which a nearly constant signal above the background level is observed. It is therefore evident that, between $T_1$ and $T_2$, the spin-state correlations experience a reduction in their dimensionality. It is interesting to note that, whereas the $l$-scans probe the correlations along the 424 ladder, the azimuthal scans shown in Figs. \ref{441}(a)-\ref{441}(h) do not correspond to a particularly high symmetry direction in reciprocal space. In fact, they are 1D reciprocal-space scans across the 441 reflection, with $\Delta(hkl)$ along the [$1x0$] direction, where $x=-b^2/a^2$ and $a$ and $b$ are the in-plane orthorhombic lattice parameters. In common, both $l$- and azimuthal scans correspond to directions in reciprocal space that are perpendicular to [110]. We conclude that, between $T_1$ and $T_2$, two-dimensional superstructure correlations take place within the (110) plane.


\subsection{Electronic transport}

Figure \ref{resistivity} displays the temperature dependence of the electrical resistance along the {\bf c}-direction $R_c(T)$. A sharp drop is seen at $T_{1}$, with a cusp-like anomaly at $T_{2}$. These anomalies are consistent with previously reported electrical measurements up to 500~K \cite{galdino2021}. No sharp anomalies in $R_c(T)$ are seen at $T_{3}$ and $T_{4}$, where the long-range spin-state order takes place. Overall, $R_c(T)$ seems to be substantially reduced between $T_{1}$ and $T_{2}$ with respect to the extrapolated behavior from lower and higher temperatures, indicating that the two-dimensional correlated phase shows substantially reduced electrical resistance along {\bf c}.

\section{Discussion}

\begin{figure}
\centering
\includegraphics[scale=0.6]{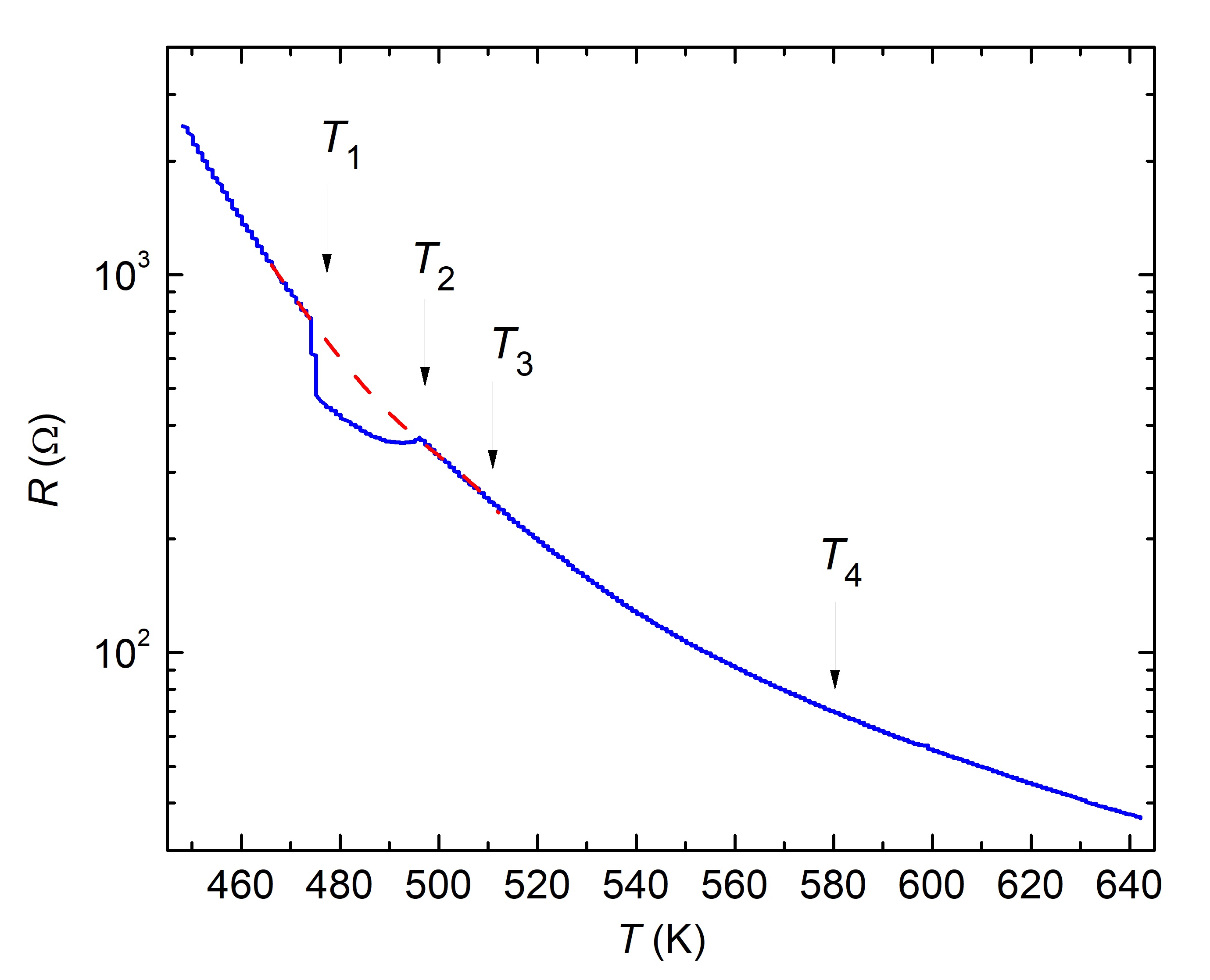}
\caption{Electrical resistance of a needle-shaped crystal of Co$_3$O$_2$BO$_3$ along the {\bf c} direction (solid line). The dashed line is the extrapolated behavior between $T_1$ and $T_2$.}
\label{resistivity}
\end{figure}

The transitions at $T_1$ and $T_2$ revealed by the $R(T)$ data [Fig. \ref{resistivity}] are also manifested in calorimetry data \cite{galdino2019magnetic} and through anomalies in the commensurate superstructure Bragg reflections (Figs. \ref{441} and \ref{reciprocal}), demonstrating these are intrinsic features of this material. On the other hand, the absence of such sharp transitions in other recent reports on the Co$_3$O$_2$BO$_3$ \cite{kazak2021spin,kazak2022} indicates that the two-dimensional correlated phase giving rise to such features is delicate and sensitive to sample growth details. Our single crystalline and powder samples of Co$_3$O$_2$BO$_3$ exhibit a $<$Co(4)-O$>$ distance of $\sim 1.95$ \AA\ at room temperature and below [Fig. \ref{distances} and ref. \cite{galdino2021}] consistent with a pure LS Co$^{3+}$ state at this site \cite{shannon1976revised}, whereas in the sample of ref. \cite{kazak2021spin} this distance is $\sim 1.98$ \AA\ that indicates a certain degree of mixing with Co$^{2+}$ or HS Co$^{3+}$ ions already at room temperature. These sample differences seem to be sufficient to preclude the formation of the fragile two-dimensional spin-state correlated state, washing out the sharp transitions at $T_1$ and $T_2$.

The temperature-dependencies of the 441 peak linewidths between $T_1$ and $T_2$ in both $l$- and azimuthal scans are intriguing [see Figs. \ref{441}(c) and \ref{fits_hk}(f)]. The linewidths are minimum (and therefore the in-plane correlation lengths are maximum) at the vicinity of $T_1$, i.e., at the frontier with the disordered phase. This is a clear indication that the transition at $T_1$ is first-order, which is also supported by the sharp discontinuity of the electrical resistance at this temperature (Fig. \ref{resistivity}). It also indicates that, at this temperature, the fraction between LS and HS Co$^{3+}$ ions is particularly favorable to stabilize a superstucture in two dimensions. Upon further warming, the ratio between HS and LS Co$^{3+}$ ions further increases, destabilizing the two-dimensional superstructure and eventually leading to three-dimensional superstructure correlations above $T_2$. At $T_3$, where the HS/LS Co$^{3+}$ occupation is presumably closer to 50\%/50\%, the three-dimensional long-range spin-state ordered state emerges and remain stable up to $T_4$, finally melting above this temperature. 

The two-dimensional superstructure found between $T_1$ and $T_2$ is very interesting and deserves further consideration. Unfortunately, the data presently at hands are insufficient to pin down unambiguously the exact nature of the electronic order associated with this phase. Nonetheless, an exciting possibility that is consistent with all our observations is suggested here. We assume that, in this temperature range, the proportion of HS Co$^{3+}$ ions is close to $x \sim 1/4$, well below $x=1/2$ demanded by the long-range spin-state ordered state displayed in Fig. \ref{superstructure}(b) with low-spin Co$^{3+}$ in site Co($4a$) and high-spin Co$^{3+}$ in site Co($4b$). Still, lattice strain may stabilize an ordered state in which spin-state-ordered 424 ladders with $x \sim 1/2$ are alternated with LS-Co$^{3+}$-rich ladders with $x \sim 0$, yielding a sample-average $x=1/4$. This configuration presumably weakens out the correlations between nearby spin-state ordered ladders, favoring a reduced dimensionality of such correlations. Upon further warming, the average proportion of HS Co$^{3+}$ ions increases towards $x=1/2$ and favors the appearance of three-dimensional spin-state correlations above $T_2$ and long-range order above $T_3$.

The drop of the electrical resistance $R(T)$ between $T_1$ and $T_2$ (see Fig. \ref{resistivity}) also deserves careful attention. First of all, the emergence of three-dimensional superstructure correlations between $T_2$ and $T_3$ and the long-range spin-state ordered phase between $T_3$ and $T_4$ are not accompanied by perceptible anomalies in the $R(T)$ curve, which rather follows the same tendency of the curve below $T_1$. Thus, only the two-dimensional superstructure state appears to be capable to impact substantially the charge transport. It is likely that, in this phase, the still incipient HS Co$^{3+}$ and the much more abundant HS Co$^{2+}$ ions, being allowed to interchange their charges while coping with spin conservation, form high-conductivity trails within the 424 3LL that lead to the observed drop in the electrical resistance. The emergence of the three-dimensional superstructure above $T_2$ possibly acts to pin down the charges back to specific sites, suppressing such a fragile channel of electrical conduction.

\section{Conclusions}

In conclusion, the homometallic Co ludwigite Co$_3$O$_2$BO$_3$ exhibits remarkable high-temperature physics. This is originated from a gradual spin-state conversion of Co$^{3+}$ that occurs simultaneously with a moderate degree of charge disorder between Co$^{2+}$ and Co$^{3+}$ above $\sim 370$ K, leading to phase transitions involving a long-range spin-state order and intermediate states at the mesoscale. Our results reveal that strongly correlated systems with mixed-valence transition-metal ions and competing spin states can have complex combined spin-state and charge correlations that are capable of generating abrupt changes in the macroscopic properties of these materials. These peculiar and unforeseen spin-charge interactions are likely mediated by lattice strain and are very sensitive to temperature. The possible sensitivity of these states to other thermodynamic variables such as pressure and magnetic field remains to be explored.

\begin{acknowledgments} 
This work was funded by FAPERJ grant E-26/210.687/2021, FAPESP Grant 2022/03539-7, and CNPq grants 305788/2020-5, 142555/2018-5 and 308607/2018-0, Brazil. Part of the research described in this paper was performed at the Canadian Light Source, a national research facility of the University of Saskatchewan, which is supported by the Canada Foundation for Innovation (CFI), the Natural Sciences and Engineering Research Council (NSERC), the National Research Council (NRC), the Canadian Institutes of Health Research (CIHR), the Government of Saskatchewan, and the University of Saskatchewan. The staff of the Brockhouse diffraction sector is ackowledged for the assistance during the X-ray diffraction experiments [Proposal GU012539]. This research also used facilities of the Brazilian Synchrotron Light Laboratory (LNLS), part of the Brazilian Center for Research in Energy and Materials (CNPEM), a private non-profit organization under the supervision of the Brazilian Ministry for Science, Technology, and Innovations (MCTI). The staff of LCTE/LNLS is acknowledged for the assistance during the transport experiment [Proposal 20221269].
\end{acknowledgments} 





\begin{thebibliography}{63}%
\makeatletter
\providecommand \@ifxundefined [1]{%
 \@ifx{#1\undefined}
}%
\providecommand \@ifnum [1]{%
 \ifnum #1\expandafter \@firstoftwo
 \else \expandafter \@secondoftwo
 \fi
}%
\providecommand \@ifx [1]{%
 \ifx #1\expandafter \@firstoftwo
 \else \expandafter \@secondoftwo
 \fi
}%
\providecommand \natexlab [1]{#1}%
\providecommand \enquote  [1]{``#1''}%
\providecommand \bibnamefont  [1]{#1}%
\providecommand \bibfnamefont [1]{#1}%
\providecommand \citenamefont [1]{#1}%
\providecommand \href@noop [0]{\@secondoftwo}%
\providecommand \href [0]{\begingroup \@sanitize@url \@href}%
\providecommand \@href[1]{\@@startlink{#1}\@@href}%
\providecommand \@@href[1]{\endgroup#1\@@endlink}%
\providecommand \@sanitize@url [0]{\catcode `\\12\catcode `\$12\catcode
  `\&12\catcode `\#12\catcode `\^12\catcode `\_12\catcode `\%12\relax}%
\providecommand \@@startlink[1]{}%
\providecommand \@@endlink[0]{}%
\providecommand \url  [0]{\begingroup\@sanitize@url \@url }%
\providecommand \@url [1]{\endgroup\@href {#1}{\urlprefix }}%
\providecommand \urlprefix  [0]{URL }%
\providecommand \Eprint [0]{\href }%
\providecommand \doibase [0]{https://doi.org/}%
\providecommand \selectlanguage [0]{\@gobble}%
\providecommand \bibinfo  [0]{\@secondoftwo}%
\providecommand \bibfield  [0]{\@secondoftwo}%
\providecommand \translation [1]{[#1]}%
\providecommand \BibitemOpen [0]{}%
\providecommand \bibitemStop [0]{}%
\providecommand \bibitemNoStop [0]{.\EOS\space}%
\providecommand \EOS [0]{\spacefactor3000\relax}%
\providecommand \BibitemShut  [1]{\csname bibitem#1\endcsname}%
\let\auto@bib@innerbib\@empty
\bibitem [{\citenamefont {Tranquada}\ \emph {et~al.}(1995)\citenamefont
  {Tranquada}, \citenamefont {Sternlieb}, \citenamefont {Axe}, \citenamefont
  {Nakamura},\ and\ \citenamefont {Uchida}}]{Tranquada1995}%
  \BibitemOpen
  \bibfield  {author} {\bibinfo {author} {\bibfnamefont {J.~M.}\ \bibnamefont
  {Tranquada}}, \bibinfo {author} {\bibfnamefont {B.~J.}\ \bibnamefont
  {Sternlieb}}, \bibinfo {author} {\bibfnamefont {J.~D.}\ \bibnamefont {Axe}},
  \bibinfo {author} {\bibfnamefont {Y.}~\bibnamefont {Nakamura}},\ and\
  \bibinfo {author} {\bibfnamefont {S.}~\bibnamefont {Uchida}},\ }\bibfield
  {title} {\bibinfo {title} {Evidence for stripe correlations of spins and
  holes in copper oxide superconductors},\ }\href
  {https://doi.org/10.1038/375561a0} {\bibfield  {journal} {\bibinfo  {journal}
  {Nature}\ }\textbf {\bibinfo {volume} {375}},\ \bibinfo {pages} {561}
  (\bibinfo {year} {1995})}\BibitemShut {NoStop}%
\bibitem [{\citenamefont {Khomskii}\ and\ \citenamefont
  {G.A.}(1997)}]{Khomskii1997}%
  \BibitemOpen
  \bibfield  {author} {\bibinfo {author} {\bibfnamefont {D.}~\bibnamefont
  {Khomskii}}\ and\ \bibinfo {author} {\bibfnamefont {S.}~\bibnamefont
  {G.A.}},\ }\bibfield  {title} {\bibinfo {title} {Interplay between spin,
  charge and orbital degrees of freedom in magnetic oxides},\ }\href
  {https://doi.org/10.1016/S0038-1098(96)00717-X} {\bibfield  {journal}
  {\bibinfo  {journal} {Solid State Communications}\ }\textbf {\bibinfo
  {volume} {102}},\ \bibinfo {pages} {87} (\bibinfo {year} {1997})},\ \bibinfo
  {note} {highlights in Condensed Matter Physics and Materials
  Science}\BibitemShut {NoStop}%
\bibitem [{\citenamefont {Salamon}\ and\ \citenamefont
  {Jaime}(2001)}]{Salamon2001}%
  \BibitemOpen
  \bibfield  {author} {\bibinfo {author} {\bibfnamefont {M.~B.}\ \bibnamefont
  {Salamon}}\ and\ \bibinfo {author} {\bibfnamefont {M.}~\bibnamefont
  {Jaime}},\ }\bibfield  {title} {\bibinfo {title} {The physics of manganites:
  Structure and transport},\ }\href {https://doi.org/10.1103/RevModPhys.73.583}
  {\bibfield  {journal} {\bibinfo  {journal} {Rev. Mod. Phys.}\ }\textbf
  {\bibinfo {volume} {73}},\ \bibinfo {pages} {583} (\bibinfo {year}
  {2001})}\BibitemShut {NoStop}%
\bibitem [{\citenamefont {Zhang}\ \emph {et~al.}(2020)\citenamefont {Zhang},
  \citenamefont {Phelan}, \citenamefont {Botana}, \citenamefont {Chen},
  \citenamefont {Zheng}, \citenamefont {Krogstad}, \citenamefont {Wang},
  \citenamefont {Qiu}, \citenamefont {Rodriguez-Rivera}, \citenamefont
  {Osborn}, \citenamefont {Rosenkranz}, \citenamefont {Norman},\ and\
  \citenamefont {Mitchell}}]{Zhang2020}%
  \BibitemOpen
  \bibfield  {author} {\bibinfo {author} {\bibfnamefont {J.}~\bibnamefont
  {Zhang}}, \bibinfo {author} {\bibfnamefont {D.}~\bibnamefont {Phelan}},
  \bibinfo {author} {\bibfnamefont {A.~S.}\ \bibnamefont {Botana}}, \bibinfo
  {author} {\bibfnamefont {Y.-S.}\ \bibnamefont {Chen}}, \bibinfo {author}
  {\bibfnamefont {H.}~\bibnamefont {Zheng}}, \bibinfo {author} {\bibfnamefont
  {M.}~\bibnamefont {Krogstad}}, \bibinfo {author} {\bibfnamefont {S.~G.}\
  \bibnamefont {Wang}}, \bibinfo {author} {\bibfnamefont {Y.}~\bibnamefont
  {Qiu}}, \bibinfo {author} {\bibfnamefont {J.~A.}\ \bibnamefont
  {Rodriguez-Rivera}}, \bibinfo {author} {\bibfnamefont {R.}~\bibnamefont
  {Osborn}}, \bibinfo {author} {\bibfnamefont {S.}~\bibnamefont {Rosenkranz}},
  \bibinfo {author} {\bibfnamefont {M.~R.}\ \bibnamefont {Norman}},\ and\
  \bibinfo {author} {\bibfnamefont {J.~F.}\ \bibnamefont {Mitchell}},\
  }\bibfield  {title} {\bibinfo {title} {Intertwined density waves in a
  metallic nickelate},\ }\href {https://doi.org/10.1038/s41467-020-19836-0}
  {\bibfield  {journal} {\bibinfo  {journal} {Nature Communications}\ }\textbf
  {\bibinfo {volume} {11}},\ \bibinfo {pages} {6003} (\bibinfo {year}
  {2020})}\BibitemShut {NoStop}%
\bibitem [{\citenamefont {Teng}\ \emph {et~al.}(2022)\citenamefont {Teng},
  \citenamefont {Chen}, \citenamefont {Ye}, \citenamefont {Rosenberg},
  \citenamefont {Liu}, \citenamefont {Yin}, \citenamefont {Jiang},
  \citenamefont {Oh}, \citenamefont {Hasan}, \citenamefont {Neubauer},
  \citenamefont {Gao}, \citenamefont {Xie}, \citenamefont {Hashimoto},
  \citenamefont {Lu}, \citenamefont {Jozwiak}, \citenamefont {Bostwick},
  \citenamefont {Rotenberg}, \citenamefont {Birgeneau}, \citenamefont {Chu},
  \citenamefont {Yi},\ and\ \citenamefont {Dai}}]{Teng2022}%
  \BibitemOpen
  \bibfield  {author} {\bibinfo {author} {\bibfnamefont {X.}~\bibnamefont
  {Teng}}, \bibinfo {author} {\bibfnamefont {L.}~\bibnamefont {Chen}}, \bibinfo
  {author} {\bibfnamefont {F.}~\bibnamefont {Ye}}, \bibinfo {author}
  {\bibfnamefont {E.}~\bibnamefont {Rosenberg}}, \bibinfo {author}
  {\bibfnamefont {Z.}~\bibnamefont {Liu}}, \bibinfo {author} {\bibfnamefont
  {J.-X.}\ \bibnamefont {Yin}}, \bibinfo {author} {\bibfnamefont {Y.-X.}\
  \bibnamefont {Jiang}}, \bibinfo {author} {\bibfnamefont {J.~S.}\ \bibnamefont
  {Oh}}, \bibinfo {author} {\bibfnamefont {M.~Z.}\ \bibnamefont {Hasan}},
  \bibinfo {author} {\bibfnamefont {K.~J.}\ \bibnamefont {Neubauer}}, \bibinfo
  {author} {\bibfnamefont {B.}~\bibnamefont {Gao}}, \bibinfo {author}
  {\bibfnamefont {Y.}~\bibnamefont {Xie}}, \bibinfo {author} {\bibfnamefont
  {M.}~\bibnamefont {Hashimoto}}, \bibinfo {author} {\bibfnamefont
  {D.}~\bibnamefont {Lu}}, \bibinfo {author} {\bibfnamefont {C.}~\bibnamefont
  {Jozwiak}}, \bibinfo {author} {\bibfnamefont {A.}~\bibnamefont {Bostwick}},
  \bibinfo {author} {\bibfnamefont {E.}~\bibnamefont {Rotenberg}}, \bibinfo
  {author} {\bibfnamefont {R.~J.}\ \bibnamefont {Birgeneau}}, \bibinfo {author}
  {\bibfnamefont {J.-H.}\ \bibnamefont {Chu}}, \bibinfo {author} {\bibfnamefont
  {M.}~\bibnamefont {Yi}},\ and\ \bibinfo {author} {\bibfnamefont
  {P.}~\bibnamefont {Dai}},\ }\bibfield  {title} {\bibinfo {title} {Discovery
  of charge density wave in a kagome lattice antiferromagnet},\ }\href
  {https://doi.org/10.1038/s41586-022-05034-z} {\bibfield  {journal} {\bibinfo
  {journal} {Nature}\ }\textbf {\bibinfo {volume} {609}},\ \bibinfo {pages}
  {490} (\bibinfo {year} {2022})}\BibitemShut {NoStop}%
\bibitem [{\citenamefont {Korotin}\ \emph {et~al.}(1996)\citenamefont
  {Korotin}, \citenamefont {Ezhov}, \citenamefont {Solovyev}, \citenamefont
  {Anisimov}, \citenamefont {Khomskii},\ and\ \citenamefont
  {Sawatzky}}]{korotin1996intermediate}%
  \BibitemOpen
  \bibfield  {author} {\bibinfo {author} {\bibfnamefont {M.}~\bibnamefont
  {Korotin}}, \bibinfo {author} {\bibfnamefont {S.~Y.}\ \bibnamefont {Ezhov}},
  \bibinfo {author} {\bibfnamefont {I.}~\bibnamefont {Solovyev}}, \bibinfo
  {author} {\bibfnamefont {V.}~\bibnamefont {Anisimov}}, \bibinfo {author}
  {\bibfnamefont {D.}~\bibnamefont {Khomskii}},\ and\ \bibinfo {author}
  {\bibfnamefont {G.}~\bibnamefont {Sawatzky}},\ }\bibfield  {title} {\bibinfo
  {title} {Intermediate-spin state and properties of \chem{LaCoO_3}},\ }\href
  {https://doi.org/10.1103/PhysRevB.54.5309} {\bibfield  {journal} {\bibinfo
  {journal} {Physical Review B}\ }\textbf {\bibinfo {volume} {54}},\ \bibinfo
  {pages} {5309} (\bibinfo {year} {1996})}\BibitemShut {NoStop}%
\bibitem [{\citenamefont {Lamonova}\ \emph {et~al.}(2011)\citenamefont
  {Lamonova}, \citenamefont {Zhitlukhina}, \citenamefont {Babkin},
  \citenamefont {Orel}, \citenamefont {Ovchinnikov},\ and\ \citenamefont
  {Pashkevich}}]{Lamonova2011}%
  \BibitemOpen
  \bibfield  {author} {\bibinfo {author} {\bibfnamefont {K.~V.}\ \bibnamefont
  {Lamonova}}, \bibinfo {author} {\bibfnamefont {E.~S.}\ \bibnamefont
  {Zhitlukhina}}, \bibinfo {author} {\bibfnamefont {R.~Y.}\ \bibnamefont
  {Babkin}}, \bibinfo {author} {\bibfnamefont {S.~M.}\ \bibnamefont {Orel}},
  \bibinfo {author} {\bibfnamefont {S.~G.}\ \bibnamefont {Ovchinnikov}},\ and\
  \bibinfo {author} {\bibfnamefont {Y.~G.}\ \bibnamefont {Pashkevich}},\
  }\bibfield  {title} {\bibinfo {title} {Intermediate-spin state of a 3d ion in
  the octahedral environment and generalization of the tanabe–sugano
  diagrams},\ }\href {https://doi.org/10.1021/jp2071265} {\bibfield  {journal}
  {\bibinfo  {journal} {The Journal of Physical Chemistry A}\ }\textbf
  {\bibinfo {volume} {115}},\ \bibinfo {pages} {13596} (\bibinfo {year}
  {2011})}\BibitemShut {NoStop}%
\bibitem [{\citenamefont {Haverkort}\ \emph {et~al.}(2006)\citenamefont
  {Haverkort}, \citenamefont {Hu}, \citenamefont {Cezar}, \citenamefont
  {Burnus}, \citenamefont {Hartmann}, \citenamefont {Reuther}, \citenamefont
  {Zobel}, \citenamefont {Lorenz}, \citenamefont {Tanaka}, \citenamefont
  {Brookes} \emph {et~al.}}]{haverkort2006spin}%
  \BibitemOpen
  \bibfield  {author} {\bibinfo {author} {\bibfnamefont {M.~W.}\ \bibnamefont
  {Haverkort}}, \bibinfo {author} {\bibfnamefont {Z.}~\bibnamefont {Hu}},
  \bibinfo {author} {\bibfnamefont {J.~C.}\ \bibnamefont {Cezar}}, \bibinfo
  {author} {\bibfnamefont {T.}~\bibnamefont {Burnus}}, \bibinfo {author}
  {\bibfnamefont {H.}~\bibnamefont {Hartmann}}, \bibinfo {author}
  {\bibfnamefont {M.}~\bibnamefont {Reuther}}, \bibinfo {author} {\bibfnamefont
  {C.}~\bibnamefont {Zobel}}, \bibinfo {author} {\bibfnamefont
  {T.}~\bibnamefont {Lorenz}}, \bibinfo {author} {\bibfnamefont
  {A.}~\bibnamefont {Tanaka}}, \bibinfo {author} {\bibfnamefont {N.~B.}\
  \bibnamefont {Brookes}}, \emph {et~al.},\ }\bibfield  {title} {\bibinfo
  {title} {Spin state transition in \chem{LaCoO_3} studied using soft
  \chem{X}-ray absorption spectroscopy and magnetic circular dichroism},\
  }\href {https://doi.org/10.1103/PhysRevLett.97.176405} {\bibfield  {journal}
  {\bibinfo  {journal} {Physical Review Letters}\ }\textbf {\bibinfo {volume}
  {97}},\ \bibinfo {pages} {176405} (\bibinfo {year} {2006})}\BibitemShut
  {NoStop}%
\bibitem [{\citenamefont {Cheng}\ \emph {et~al.}(2012)\citenamefont {Cheng},
  \citenamefont {Zhou}, \citenamefont {Hu}, \citenamefont {Suchomel},
  \citenamefont {Chin}, \citenamefont {Kuo}, \citenamefont {Lin}, \citenamefont
  {Chen}, \citenamefont {Pi}, \citenamefont {Chen} \emph
  {et~al.}}]{cheng2012spin}%
  \BibitemOpen
  \bibfield  {author} {\bibinfo {author} {\bibfnamefont {J.-G.}\ \bibnamefont
  {Cheng}}, \bibinfo {author} {\bibfnamefont {J.-S.}\ \bibnamefont {Zhou}},
  \bibinfo {author} {\bibfnamefont {Z.}~\bibnamefont {Hu}}, \bibinfo {author}
  {\bibfnamefont {M.}~\bibnamefont {Suchomel}}, \bibinfo {author}
  {\bibfnamefont {Y.}~\bibnamefont {Chin}}, \bibinfo {author} {\bibfnamefont
  {C.}~\bibnamefont {Kuo}}, \bibinfo {author} {\bibfnamefont {H.-J.}\
  \bibnamefont {Lin}}, \bibinfo {author} {\bibfnamefont {J.}~\bibnamefont
  {Chen}}, \bibinfo {author} {\bibfnamefont {D.}~\bibnamefont {Pi}}, \bibinfo
  {author} {\bibfnamefont {C.}~\bibnamefont {Chen}}, \emph {et~al.},\
  }\bibfield  {title} {\bibinfo {title} {Spin-state transition in
  \chem{Ba_2Co_9O_{14}}},\ }\href {https://doi.org/10.1103/PhysRevB.85.094424}
  {\bibfield  {journal} {\bibinfo  {journal} {Physical Review B}\ }\textbf
  {\bibinfo {volume} {85}},\ \bibinfo {pages} {094424} (\bibinfo {year}
  {2012})}\BibitemShut {NoStop}%
\bibitem [{\citenamefont {Kazak}\ \emph {et~al.}(2021)\citenamefont {Kazak},
  \citenamefont {Platunov}, \citenamefont {Knyazev}, \citenamefont {Molokeev},
  \citenamefont {Gorev}, \citenamefont {Ovchinnikov}, \citenamefont
  {Pchelkina}, \citenamefont {Gapontsev}, \citenamefont {Streltsov},
  \citenamefont {Bartolom{\'e}} \emph {et~al.}}]{kazak2021spin}%
  \BibitemOpen
  \bibfield  {author} {\bibinfo {author} {\bibfnamefont {N.}~\bibnamefont
  {Kazak}}, \bibinfo {author} {\bibfnamefont {M.}~\bibnamefont {Platunov}},
  \bibinfo {author} {\bibfnamefont {Y.~V.}\ \bibnamefont {Knyazev}}, \bibinfo
  {author} {\bibfnamefont {M.}~\bibnamefont {Molokeev}}, \bibinfo {author}
  {\bibfnamefont {M.}~\bibnamefont {Gorev}}, \bibinfo {author} {\bibfnamefont
  {S.}~\bibnamefont {Ovchinnikov}}, \bibinfo {author} {\bibfnamefont
  {Z.}~\bibnamefont {Pchelkina}}, \bibinfo {author} {\bibfnamefont
  {V.}~\bibnamefont {Gapontsev}}, \bibinfo {author} {\bibfnamefont
  {S.}~\bibnamefont {Streltsov}}, \bibinfo {author} {\bibfnamefont
  {J.}~\bibnamefont {Bartolom{\'e}}}, \emph {et~al.},\ }\bibfield  {title}
  {\bibinfo {title} {Spin state crossover in \chem{Co_3BO_5}},\ }\href
  {https://doi.org/10.1103/PhysRevB.103.094445} {\bibfield  {journal} {\bibinfo
   {journal} {Physical Review B}\ }\textbf {\bibinfo {volume} {103}},\ \bibinfo
  {pages} {094445} (\bibinfo {year} {2021})}\BibitemShut {NoStop}%
\bibitem [{\citenamefont {Doumerc}\ \emph {et~al.}(1999)\citenamefont
  {Doumerc}, \citenamefont {Coutanceau}, \citenamefont {Fournés},
  \citenamefont {Grenier}, \citenamefont {Pouchard},\ and\ \citenamefont
  {Wattiaux}}]{Doumerc1999}%
  \BibitemOpen
  \bibfield  {author} {\bibinfo {author} {\bibfnamefont {J.-P.}\ \bibnamefont
  {Doumerc}}, \bibinfo {author} {\bibfnamefont {M.}~\bibnamefont {Coutanceau}},
  \bibinfo {author} {\bibfnamefont {L.}~\bibnamefont {Fournés}}, \bibinfo
  {author} {\bibfnamefont {J.-C.}\ \bibnamefont {Grenier}}, \bibinfo {author}
  {\bibfnamefont {M.}~\bibnamefont {Pouchard}},\ and\ \bibinfo {author}
  {\bibfnamefont {A.}~\bibnamefont {Wattiaux}},\ }\bibfield  {title} {\bibinfo
  {title} {Mössbauer investigation of \chem{^{57}Fe}-doped
  \chem{TlSr_2CoO_5}},\ }\href {https://doi.org/10.1016/S1387-1609(00)88577-1}
  {\bibfield  {journal} {\bibinfo  {journal} {Comptes Rendus de l'Académie des
  Sciences - Series IIC - Chemistry}\ }\textbf {\bibinfo {volume} {2}},\
  \bibinfo {pages} {637} (\bibinfo {year} {1999})}\BibitemShut {NoStop}%
\bibitem [{\citenamefont {Doumerc}\ \emph {et~al.}(2001)\citenamefont
  {Doumerc}, \citenamefont {Coutanceau}, \citenamefont {Demourgues},
  \citenamefont {Elkaim}, \citenamefont {Grenier},\ and\ \citenamefont
  {Pouchard}}]{Doumerc2001}%
  \BibitemOpen
  \bibfield  {author} {\bibinfo {author} {\bibfnamefont {J.-P.}\ \bibnamefont
  {Doumerc}}, \bibinfo {author} {\bibfnamefont {M.}~\bibnamefont {Coutanceau}},
  \bibinfo {author} {\bibfnamefont {A.}~\bibnamefont {Demourgues}}, \bibinfo
  {author} {\bibfnamefont {E.}~\bibnamefont {Elkaim}}, \bibinfo {author}
  {\bibfnamefont {J.-C.}\ \bibnamefont {Grenier}},\ and\ \bibinfo {author}
  {\bibfnamefont {M.}~\bibnamefont {Pouchard}},\ }\bibfield  {title} {\bibinfo
  {title} {Crystal structure of the thallium strontium cobaltite
  \chem{TlSr_2CoO_5} and its relationship to the electronic properties},\
  }\href {https://doi.org/10.1039/B003198O} {\bibfield  {journal} {\bibinfo
  {journal} {J. Mater. Chem.}\ }\textbf {\bibinfo {volume} {11}},\ \bibinfo
  {pages} {78} (\bibinfo {year} {2001})}\BibitemShut {NoStop}%
\bibitem [{\citenamefont {Khomskii}\ and\ \citenamefont
  {L{\"o}w}(2004)}]{khomskii2004superstructures}%
  \BibitemOpen
  \bibfield  {author} {\bibinfo {author} {\bibfnamefont {D.}~\bibnamefont
  {Khomskii}}\ and\ \bibinfo {author} {\bibfnamefont {U.}~\bibnamefont
  {L{\"o}w}},\ }\bibfield  {title} {\bibinfo {title} {Superstructures at low
  spin-high spin transitions},\ }\href
  {https://doi.org/10.1103/PhysRevB.69.184401} {\bibfield  {journal} {\bibinfo
  {journal} {Physical Review B}\ }\textbf {\bibinfo {volume} {69}},\ \bibinfo
  {pages} {184401} (\bibinfo {year} {2004})}\BibitemShut {NoStop}%
\bibitem [{\citenamefont {Collet}\ \emph {et~al.}(2012)\citenamefont {Collet},
  \citenamefont {Watanabe}, \citenamefont {Br\'efuel}, \citenamefont
  {Palatinus}, \citenamefont {Roudaut}, \citenamefont {Toupet}, \citenamefont
  {Tanaka}, \citenamefont {Tuchagues}, \citenamefont {Fertey}, \citenamefont
  {Ravy}, \citenamefont {Toudic},\ and\ \citenamefont {Cailleau}}]{Collet2012}%
  \BibitemOpen
  \bibfield  {author} {\bibinfo {author} {\bibfnamefont {E.}~\bibnamefont
  {Collet}}, \bibinfo {author} {\bibfnamefont {H.}~\bibnamefont {Watanabe}},
  \bibinfo {author} {\bibfnamefont {N.}~\bibnamefont {Br\'efuel}}, \bibinfo
  {author} {\bibfnamefont {L.}~\bibnamefont {Palatinus}}, \bibinfo {author}
  {\bibfnamefont {L.}~\bibnamefont {Roudaut}}, \bibinfo {author} {\bibfnamefont
  {L.}~\bibnamefont {Toupet}}, \bibinfo {author} {\bibfnamefont
  {K.}~\bibnamefont {Tanaka}}, \bibinfo {author} {\bibfnamefont {J.-P.}\
  \bibnamefont {Tuchagues}}, \bibinfo {author} {\bibfnamefont {P.}~\bibnamefont
  {Fertey}}, \bibinfo {author} {\bibfnamefont {S.}~\bibnamefont {Ravy}},
  \bibinfo {author} {\bibfnamefont {B.}~\bibnamefont {Toudic}},\ and\ \bibinfo
  {author} {\bibfnamefont {H.}~\bibnamefont {Cailleau}},\ }\bibfield  {title}
  {\bibinfo {title} {Aperiodic spin state ordering of bistable molecules and
  its photoinduced erasing},\ }\href
  {https://doi.org/10.1103/PhysRevLett.109.257206} {\bibfield  {journal}
  {\bibinfo  {journal} {Phys. Rev. Lett.}\ }\textbf {\bibinfo {volume} {109}},\
  \bibinfo {pages} {257206} (\bibinfo {year} {2012})}\BibitemShut {NoStop}%
\bibitem [{\citenamefont {Hu}\ \emph {et~al.}(2012)\citenamefont {Hu},
  \citenamefont {Wu}, \citenamefont {Koethe}, \citenamefont {Barilo},
  \citenamefont {Shiryaev}, \citenamefont {Bychkov}, \citenamefont
  {Schüßler-Langeheine}, \citenamefont {Lorenz}, \citenamefont {Tanaka},
  \citenamefont {Hsieh}, \citenamefont {Lin}, \citenamefont {Chen},
  \citenamefont {Brookes}, \citenamefont {Agrestini}, \citenamefont {Chin},
  \citenamefont {Rotter},\ and\ \citenamefont {Tjeng}}]{Hu_2012}%
  \BibitemOpen
  \bibfield  {author} {\bibinfo {author} {\bibfnamefont {Z.}~\bibnamefont
  {Hu}}, \bibinfo {author} {\bibfnamefont {H.}~\bibnamefont {Wu}}, \bibinfo
  {author} {\bibfnamefont {T.~C.}\ \bibnamefont {Koethe}}, \bibinfo {author}
  {\bibfnamefont {S.~N.}\ \bibnamefont {Barilo}}, \bibinfo {author}
  {\bibfnamefont {S.~V.}\ \bibnamefont {Shiryaev}}, \bibinfo {author}
  {\bibfnamefont {G.~L.}\ \bibnamefont {Bychkov}}, \bibinfo {author}
  {\bibfnamefont {C.}~\bibnamefont {Schüßler-Langeheine}}, \bibinfo {author}
  {\bibfnamefont {T.}~\bibnamefont {Lorenz}}, \bibinfo {author} {\bibfnamefont
  {A.}~\bibnamefont {Tanaka}}, \bibinfo {author} {\bibfnamefont {H.~H.}\
  \bibnamefont {Hsieh}}, \bibinfo {author} {\bibfnamefont {H.-J.}\ \bibnamefont
  {Lin}}, \bibinfo {author} {\bibfnamefont {C.~T.}\ \bibnamefont {Chen}},
  \bibinfo {author} {\bibfnamefont {N.~B.}\ \bibnamefont {Brookes}}, \bibinfo
  {author} {\bibfnamefont {S.}~\bibnamefont {Agrestini}}, \bibinfo {author}
  {\bibfnamefont {Y.-Y.}\ \bibnamefont {Chin}}, \bibinfo {author}
  {\bibfnamefont {M.}~\bibnamefont {Rotter}},\ and\ \bibinfo {author}
  {\bibfnamefont {L.~H.}\ \bibnamefont {Tjeng}},\ }\bibfield  {title} {\bibinfo
  {title} {Spin-state order/disorder and metal–insulator transition in
  \chem{GdBaCo_2O_{5.5}}: experimental determination of the underlying
  electronic structure},\ }\href
  {https://doi.org/10.1088/1367-2630/14/12/123025} {\bibfield  {journal}
  {\bibinfo  {journal} {New Journal of Physics}\ }\textbf {\bibinfo {volume}
  {14}},\ \bibinfo {pages} {123025} (\bibinfo {year} {2012})}\BibitemShut
  {NoStop}%
\bibitem [{\citenamefont {Ikeda}\ \emph {et~al.}(2016)\citenamefont {Ikeda},
  \citenamefont {Nomura}, \citenamefont {Matsuda}, \citenamefont {Matsuo},
  \citenamefont {Kindo},\ and\ \citenamefont {Sato}}]{Ikeda2016}%
  \BibitemOpen
  \bibfield  {author} {\bibinfo {author} {\bibfnamefont {A.}~\bibnamefont
  {Ikeda}}, \bibinfo {author} {\bibfnamefont {T.}~\bibnamefont {Nomura}},
  \bibinfo {author} {\bibfnamefont {Y.~H.}\ \bibnamefont {Matsuda}}, \bibinfo
  {author} {\bibfnamefont {A.}~\bibnamefont {Matsuo}}, \bibinfo {author}
  {\bibfnamefont {K.}~\bibnamefont {Kindo}},\ and\ \bibinfo {author}
  {\bibfnamefont {K.}~\bibnamefont {Sato}},\ }\bibfield  {title} {\bibinfo
  {title} {Spin state ordering of strongly correlating \chem{LaCoO_3} induced
  at ultrahigh magnetic fields},\ }\href
  {https://doi.org/10.1103/PhysRevB.93.220401} {\bibfield  {journal} {\bibinfo
  {journal} {Phys. Rev. B}\ }\textbf {\bibinfo {volume} {93}},\ \bibinfo
  {pages} {220401} (\bibinfo {year} {2016})}\BibitemShut {NoStop}%
\bibitem [{\citenamefont {Murnaghan}\ \emph {et~al.}(2014)\citenamefont
  {Murnaghan}, \citenamefont {Carbonera}, \citenamefont {Toupet}, \citenamefont
  {Griffin}, \citenamefont {Dîrtu}, \citenamefont {Desplanches}, \citenamefont
  {Garcia}, \citenamefont {Collet}, \citenamefont {Létard},\ and\
  \citenamefont {Morgan}}]{Murnaghan2014}%
  \BibitemOpen
  \bibfield  {author} {\bibinfo {author} {\bibfnamefont {K.}~\bibnamefont
  {Murnaghan}}, \bibinfo {author} {\bibfnamefont {C.}~\bibnamefont
  {Carbonera}}, \bibinfo {author} {\bibfnamefont {L.}~\bibnamefont {Toupet}},
  \bibinfo {author} {\bibfnamefont {M.}~\bibnamefont {Griffin}}, \bibinfo
  {author} {\bibfnamefont {M.}~\bibnamefont {Dîrtu}}, \bibinfo {author}
  {\bibfnamefont {C.}~\bibnamefont {Desplanches}}, \bibinfo {author}
  {\bibfnamefont {Y.}~\bibnamefont {Garcia}}, \bibinfo {author} {\bibfnamefont
  {E.}~\bibnamefont {Collet}}, \bibinfo {author} {\bibfnamefont
  {J.}~\bibnamefont {Létard}},\ and\ \bibinfo {author} {\bibfnamefont
  {G.}~\bibnamefont {Morgan}},\ }\bibfield  {title} {\bibinfo {title}
  {Spin-state ordering on one sub-lattice of a mononuclear iron(\chem{III})
  spin crossover complex exhibiting \chem{LIESST} and \chem{TIESST}},\ }\href
  {https://doi.org/10.1002/chem.201400286} {\bibfield  {journal} {\bibinfo
  {journal} {Chemistry}\ }\textbf {\bibinfo {volume} {20}},\ \bibinfo {pages}
  {5613} (\bibinfo {year} {2014})}\BibitemShut {NoStop}%
\bibitem [{\citenamefont {Takeuchi}\ \emph {et~al.}(1950)\citenamefont
  {Takeuchi}, \citenamefont {Watanabe},\ and\ \citenamefont
  {Ito}}]{takeuchi1950crystal}%
  \BibitemOpen
  \bibfield  {author} {\bibinfo {author} {\bibfnamefont {Y.}~\bibnamefont
  {Takeuchi}}, \bibinfo {author} {\bibfnamefont {T.}~\bibnamefont {Watanabe}},\
  and\ \bibinfo {author} {\bibfnamefont {T.}~\bibnamefont {Ito}},\ }\bibfield
  {title} {\bibinfo {title} {The crystal structures of warwickite, ludwigite
  and pinakiolite},\ }\href {https://doi.org/10.1107/S0365110X50000252}
  {\bibfield  {journal} {\bibinfo  {journal} {Acta Crystallographica}\ }\textbf
  {\bibinfo {volume} {3}},\ \bibinfo {pages} {98} (\bibinfo {year}
  {1950})}\BibitemShut {NoStop}%
\bibitem [{\citenamefont {Wiedenmann}\ and\ \citenamefont
  {Burlet}(1978)}]{wiedenmann1978magnetic}%
  \BibitemOpen
  \bibfield  {author} {\bibinfo {author} {\bibfnamefont {A.}~\bibnamefont
  {Wiedenmann}}\ and\ \bibinfo {author} {\bibfnamefont {P.}~\bibnamefont
  {Burlet}},\ }\bibfield  {title} {\bibinfo {title} {Magnetic behavior of
  imperfect quasi one dimensional insulators \chem{FeMgBO_4} and
  \chem{FeMg_2BO_5}: Spin glass systems?},\ }\href
  {https://doi.org/10.1051/jphyscol:19786320} {\bibfield  {journal} {\bibinfo
  {journal} {Le Journal de Physique Colloques}\ }\textbf {\bibinfo {volume}
  {39}},\ \bibinfo {pages} {C6} (\bibinfo {year} {1978})}\BibitemShut {NoStop}%
\bibitem [{\citenamefont {Norrestam}\ \emph {et~al.}(1994)\citenamefont
  {Norrestam}, \citenamefont {Kritikos}, \citenamefont {Nielsen}, \citenamefont
  {S{\o}tofte},\ and\ \citenamefont {Thorup}}]{norrestam1994structural}%
  \BibitemOpen
  \bibfield  {author} {\bibinfo {author} {\bibfnamefont {R.}~\bibnamefont
  {Norrestam}}, \bibinfo {author} {\bibfnamefont {M.}~\bibnamefont {Kritikos}},
  \bibinfo {author} {\bibfnamefont {K.}~\bibnamefont {Nielsen}}, \bibinfo
  {author} {\bibfnamefont {I.}~\bibnamefont {S{\o}tofte}},\ and\ \bibinfo
  {author} {\bibfnamefont {N.}~\bibnamefont {Thorup}},\ }\bibfield  {title}
  {\bibinfo {title} {Structural characterizations of two synthetic
  \chem{Ni}-ludwigites, and some semiempirical \chem{EHTB} calculations on the
  ludwigite structure type},\ }\href {https://doi.org/10.1006/jssc.1994.1220}
  {\bibfield  {journal} {\bibinfo  {journal} {Journal of Solid State
  Chemistry}\ }\textbf {\bibinfo {volume} {111}},\ \bibinfo {pages} {217}
  (\bibinfo {year} {1994})}\BibitemShut {NoStop}%
\bibitem [{\citenamefont {Fernandes}\ \emph {et~al.}(1998)\citenamefont
  {Fernandes}, \citenamefont {Guimar{\~a}es}, \citenamefont {Continentino},
  \citenamefont {Borges}, \citenamefont {Sulpice}, \citenamefont {Tholence},
  \citenamefont {Siqueira}, \citenamefont {Zawislak}, \citenamefont
  {da~Cunha},\ and\ \citenamefont {dos Santos}}]{fernandes1998magnetic}%
  \BibitemOpen
  \bibfield  {author} {\bibinfo {author} {\bibfnamefont {J.~C.}\ \bibnamefont
  {Fernandes}}, \bibinfo {author} {\bibfnamefont {R.~B.}\ \bibnamefont
  {Guimar{\~a}es}}, \bibinfo {author} {\bibfnamefont {M.~A.}\ \bibnamefont
  {Continentino}}, \bibinfo {author} {\bibfnamefont {H.~A.}\ \bibnamefont
  {Borges}}, \bibinfo {author} {\bibfnamefont {A.}~\bibnamefont {Sulpice}},
  \bibinfo {author} {\bibfnamefont {J.~L.}\ \bibnamefont {Tholence}}, \bibinfo
  {author} {\bibfnamefont {J.~L.}\ \bibnamefont {Siqueira}}, \bibinfo {author}
  {\bibfnamefont {L.~I.}\ \bibnamefont {Zawislak}}, \bibinfo {author}
  {\bibfnamefont {J.~B.~M.}\ \bibnamefont {da~Cunha}},\ and\ \bibinfo {author}
  {\bibfnamefont {C.~A.}\ \bibnamefont {dos Santos}},\ }\bibfield  {title}
  {\bibinfo {title} {Magnetic interactions in the ludwigite
  \chem{Ni_2FeO_2BO_3}},\ }\href {https://doi.org/10.1103/PhysRevB.58.287}
  {\bibfield  {journal} {\bibinfo  {journal} {Physical Review B}\ }\textbf
  {\bibinfo {volume} {58}},\ \bibinfo {pages} {287} (\bibinfo {year}
  {1998})}\BibitemShut {NoStop}%
\bibitem [{\citenamefont {Irwin}\ and\ \citenamefont
  {Peterson}(1999)}]{irwin1999crystal}%
  \BibitemOpen
  \bibfield  {author} {\bibinfo {author} {\bibfnamefont {M.~B.}\ \bibnamefont
  {Irwin}}\ and\ \bibinfo {author} {\bibfnamefont {R.~C.}\ \bibnamefont
  {Peterson}},\ }\bibfield  {title} {\bibinfo {title} {The crystal structure of
  ludwigite},\ }\href@noop {} {\bibfield  {journal} {\bibinfo  {journal}
  {Canadian mineralogist}\ }\textbf {\bibinfo {volume} {37}},\ \bibinfo {pages}
  {939} (\bibinfo {year} {1999})}\BibitemShut {NoStop}%
\bibitem [{\citenamefont {Guimar{\~a}es}\ \emph {et~al.}(1999)\citenamefont
  {Guimar{\~a}es}, \citenamefont {Mir}, \citenamefont {Fernandes},
  \citenamefont {Continentino}, \citenamefont {Borges}, \citenamefont
  {Cernicchiaro}, \citenamefont {Fontes}, \citenamefont {Candela},\ and\
  \citenamefont {Baggio-Saitovich}}]{guimaraes1999cation}%
  \BibitemOpen
  \bibfield  {author} {\bibinfo {author} {\bibfnamefont {R.~B.}\ \bibnamefont
  {Guimar{\~a}es}}, \bibinfo {author} {\bibfnamefont {M.}~\bibnamefont {Mir}},
  \bibinfo {author} {\bibfnamefont {J.~C.}\ \bibnamefont {Fernandes}}, \bibinfo
  {author} {\bibfnamefont {M.~A.}\ \bibnamefont {Continentino}}, \bibinfo
  {author} {\bibfnamefont {H.~A.}\ \bibnamefont {Borges}}, \bibinfo {author}
  {\bibfnamefont {G.}~\bibnamefont {Cernicchiaro}}, \bibinfo {author}
  {\bibfnamefont {M.~B.}\ \bibnamefont {Fontes}}, \bibinfo {author}
  {\bibfnamefont {D.~R.~S.}\ \bibnamefont {Candela}},\ and\ \bibinfo {author}
  {\bibfnamefont {E.}~\bibnamefont {Baggio-Saitovich}},\ }\bibfield  {title}
  {\bibinfo {title} {Cation-mediated interaction and weak ferromagnetism in
  \chem{Fe_3O_2BO_3}},\ }\href {https://doi.org/10.1103/PhysRevB.60.6617}
  {\bibfield  {journal} {\bibinfo  {journal} {Physical Review B}\ }\textbf
  {\bibinfo {volume} {60}},\ \bibinfo {pages} {6617} (\bibinfo {year}
  {1999})}\BibitemShut {NoStop}%
\bibitem [{\citenamefont {Continentino}\ \emph {et~al.}(1999)\citenamefont
  {Continentino}, \citenamefont {Fernandes}, \citenamefont {Borges},
  \citenamefont {Sulpice}, \citenamefont {Tholence}, \citenamefont {Siqueira},
  \citenamefont {da~Cunha}, \citenamefont {dos Santos} \emph
  {et~al.}}]{continentino1999magnetic}%
  \BibitemOpen
  \bibfield  {author} {\bibinfo {author} {\bibfnamefont {M.}~\bibnamefont
  {Continentino}}, \bibinfo {author} {\bibfnamefont {J.}~\bibnamefont
  {Fernandes}}, \bibinfo {author} {\bibfnamefont {H.}~\bibnamefont {Borges}},
  \bibinfo {author} {\bibfnamefont {A.}~\bibnamefont {Sulpice}}, \bibinfo
  {author} {\bibfnamefont {J.-L.}\ \bibnamefont {Tholence}}, \bibinfo {author}
  {\bibfnamefont {J.}~\bibnamefont {Siqueira}}, \bibinfo {author}
  {\bibfnamefont {J.}~\bibnamefont {da~Cunha}}, \bibinfo {author}
  {\bibfnamefont {C.}~\bibnamefont {dos Santos}}, \emph {et~al.},\ }\bibfield
  {title} {\bibinfo {title} {Magnetic interactions in the monoclinic ludwigite
  \chem{Cu_2FeO_2BO_3}},\ }\href {https://doi.org/10.1007/s100510050805}
  {\bibfield  {journal} {\bibinfo  {journal} {The European Physical Journal
  B-Condensed Matter and Complex Systems}\ }\textbf {\bibinfo {volume} {9}},\
  \bibinfo {pages} {613} (\bibinfo {year} {1999})}\BibitemShut {NoStop}%
\bibitem [{\citenamefont {Fernandes}\ \emph {et~al.}(2000)\citenamefont
  {Fernandes}, \citenamefont {Guimar{\~a}es}, \citenamefont {Continentino},
  \citenamefont {Ghivelder},\ and\ \citenamefont
  {Freitas}}]{fernandes2000specific}%
  \BibitemOpen
  \bibfield  {author} {\bibinfo {author} {\bibfnamefont {J.~C.}\ \bibnamefont
  {Fernandes}}, \bibinfo {author} {\bibfnamefont {R.~B.}\ \bibnamefont
  {Guimar{\~a}es}}, \bibinfo {author} {\bibfnamefont {M.~A.}\ \bibnamefont
  {Continentino}}, \bibinfo {author} {\bibfnamefont {L.}~\bibnamefont
  {Ghivelder}},\ and\ \bibinfo {author} {\bibfnamefont {R.~S.}\ \bibnamefont
  {Freitas}},\ }\bibfield  {title} {\bibinfo {title} {Specific heat of
  \chem{Fe_3O_2BO_3}: \chem{E}vidence for a \chem{W}igner glass phase},\ }\href
  {https://doi.org/10.1103/PhysRevB.61.R850} {\bibfield  {journal} {\bibinfo
  {journal} {Physical Review B}\ }\textbf {\bibinfo {volume} {61}},\ \bibinfo
  {pages} {R850} (\bibinfo {year} {2000})}\BibitemShut {NoStop}%
\bibitem [{\citenamefont {Mir}\ \emph {et~al.}(2001)\citenamefont {Mir},
  \citenamefont {Guimaraes}, \citenamefont {Fernandes}, \citenamefont
  {Continentino}, \citenamefont {Doriguetto}, \citenamefont {Mascarenhas},
  \citenamefont {Ellena}, \citenamefont {Castellano}, \citenamefont {Freitas},\
  and\ \citenamefont {Ghivelder}}]{mir2001structural}%
  \BibitemOpen
  \bibfield  {author} {\bibinfo {author} {\bibfnamefont {M.}~\bibnamefont
  {Mir}}, \bibinfo {author} {\bibfnamefont {R.~B.}\ \bibnamefont {Guimaraes}},
  \bibinfo {author} {\bibfnamefont {J.~C.}\ \bibnamefont {Fernandes}}, \bibinfo
  {author} {\bibfnamefont {M.~A.}\ \bibnamefont {Continentino}}, \bibinfo
  {author} {\bibfnamefont {A.~C.}\ \bibnamefont {Doriguetto}}, \bibinfo
  {author} {\bibfnamefont {Y.~P.}\ \bibnamefont {Mascarenhas}}, \bibinfo
  {author} {\bibfnamefont {J.}~\bibnamefont {Ellena}}, \bibinfo {author}
  {\bibfnamefont {E.~E.}\ \bibnamefont {Castellano}}, \bibinfo {author}
  {\bibfnamefont {R.~S.}\ \bibnamefont {Freitas}},\ and\ \bibinfo {author}
  {\bibfnamefont {L.}~\bibnamefont {Ghivelder}},\ }\bibfield  {title} {\bibinfo
  {title} {Structural transition and pair formation in \chem{Fe_3O_2BO_3}},\
  }\href {https://doi.org/10.1103/PhysRevLett.87.147201} {\bibfield  {journal}
  {\bibinfo  {journal} {Physical Review Letters}\ }\textbf {\bibinfo {volume}
  {87}},\ \bibinfo {pages} {147201} (\bibinfo {year} {2001})}\BibitemShut
  {NoStop}%
\bibitem [{\citenamefont {Continentino}\ \emph {et~al.}(2001)\citenamefont
  {Continentino}, \citenamefont {Boechat}, \citenamefont {Guimaraes},
  \citenamefont {Fernandes},\ and\ \citenamefont
  {Ghivelder}}]{continentino2001magnetic}%
  \BibitemOpen
  \bibfield  {author} {\bibinfo {author} {\bibfnamefont {M.}~\bibnamefont
  {Continentino}}, \bibinfo {author} {\bibfnamefont {B.}~\bibnamefont
  {Boechat}}, \bibinfo {author} {\bibfnamefont {R.}~\bibnamefont {Guimaraes}},
  \bibinfo {author} {\bibfnamefont {J.}~\bibnamefont {Fernandes}},\ and\
  \bibinfo {author} {\bibfnamefont {L.}~\bibnamefont {Ghivelder}},\ }\bibfield
  {title} {\bibinfo {title} {Magnetic and transport properties of
  low-dimensional oxi-borates},\ }\href
  {https://doi.org/10.1016/S0304-8853(00)00927-6} {\bibfield  {journal}
  {\bibinfo  {journal} {Journal of magnetism and magnetic materials}\ }\textbf
  {\bibinfo {volume} {226}},\ \bibinfo {pages} {427} (\bibinfo {year}
  {2001})}\BibitemShut {NoStop}%
\bibitem [{\citenamefont {Whangbo}\ \emph {et~al.}(2002)\citenamefont
  {Whangbo}, \citenamefont {Koo}, \citenamefont {Dumas},\ and\ \citenamefont
  {Continentino}}]{whangbo2002theoretical}%
  \BibitemOpen
  \bibfield  {author} {\bibinfo {author} {\bibfnamefont {M.-H.}\ \bibnamefont
  {Whangbo}}, \bibinfo {author} {\bibfnamefont {H.-J.}\ \bibnamefont {Koo}},
  \bibinfo {author} {\bibfnamefont {J.}~\bibnamefont {Dumas}},\ and\ \bibinfo
  {author} {\bibfnamefont {M.}~\bibnamefont {Continentino}},\ }\bibfield
  {title} {\bibinfo {title} {Theoretical investigation of the spin exchange
  interactions and magnetic properties of the homometallic ludwigite
  \chem{Fe_3O_2BO_3}},\ }\href {https://doi.org/10.1021/ic010956q} {\bibfield
  {journal} {\bibinfo  {journal} {Inorganic chemistry}\ }\textbf {\bibinfo
  {volume} {41}},\ \bibinfo {pages} {2193} (\bibinfo {year}
  {2002})}\BibitemShut {NoStop}%
\bibitem [{\citenamefont {Larrea~J.}\ \emph {et~al.}(2004)\citenamefont
  {Larrea~J.}, \citenamefont {S\'anchez}, \citenamefont {Litterst},
  \citenamefont {Baggio-Saitovitch}, \citenamefont {Fernandes}, \citenamefont
  {Guimar\~aes},\ and\ \citenamefont {Continentino}}]{sanchez2004magnetism}%
  \BibitemOpen
  \bibfield  {author} {\bibinfo {author} {\bibfnamefont {J.}~\bibnamefont
  {Larrea~J.}}, \bibinfo {author} {\bibfnamefont {D.~R.}\ \bibnamefont
  {S\'anchez}}, \bibinfo {author} {\bibfnamefont {F.~J.}\ \bibnamefont
  {Litterst}}, \bibinfo {author} {\bibfnamefont {E.~M.}\ \bibnamefont
  {Baggio-Saitovitch}}, \bibinfo {author} {\bibfnamefont {J.~C.}\ \bibnamefont
  {Fernandes}}, \bibinfo {author} {\bibfnamefont {R.~B.}\ \bibnamefont
  {Guimar\~aes}},\ and\ \bibinfo {author} {\bibfnamefont {M.~A.}\ \bibnamefont
  {Continentino}},\ }\bibfield  {title} {\bibinfo {title} {Magnetism and charge
  ordering in \chem{Fe_3O_2BO_3} studied by \chem{^{57}Fe} \chem{M}\"ossbauer
  spectroscopy},\ }\href {https://doi.org/10.1103/PhysRevB.70.174452}
  {\bibfield  {journal} {\bibinfo  {journal} {Phys. Rev. B}\ }\textbf {\bibinfo
  {volume} {70}},\ \bibinfo {pages} {174452} (\bibinfo {year}
  {2004})}\BibitemShut {NoStop}%
\bibitem [{\citenamefont {Continentino}\ \emph {et~al.}(2005)\citenamefont
  {Continentino}, \citenamefont {Fernandes}, \citenamefont {Guimarães},
  \citenamefont {Boechat},\ and\ \citenamefont
  {Saguia}}]{borates2005frontiers}%
  \BibitemOpen
  \bibfield  {author} {\bibinfo {author} {\bibfnamefont {M.~A.}\ \bibnamefont
  {Continentino}}, \bibinfo {author} {\bibfnamefont {J.~C.}\ \bibnamefont
  {Fernandes}}, \bibinfo {author} {\bibfnamefont {R.~B.}\ \bibnamefont
  {Guimarães}}, \bibinfo {author} {\bibfnamefont {B.}~\bibnamefont
  {Boechat}},\ and\ \bibinfo {author} {\bibfnamefont {A.}~\bibnamefont
  {Saguia}},\ }\bibfield  {title} {\bibinfo {title} {Magnetism in highly
  anisotropic borates: Experiment and theory},\ }in\ \href
  {https://doi.org/10.1007/3-540-27284-4_14} {\emph {\bibinfo {booktitle}
  {Frontiers in Magnetic Materials}}}\ (\bibinfo  {publisher} {Springer},\
  \bibinfo {year} {2005})\ p.\ \bibinfo {pages} {385}\BibitemShut {NoStop}%
\bibitem [{\citenamefont {Fernandes}\ \emph {et~al.}(2005)\citenamefont
  {Fernandes}, \citenamefont {Guimar{\~a}es}, \citenamefont {Continentino},
  \citenamefont {Ziemath}, \citenamefont {Walmsley}, \citenamefont
  {Monteverde}, \citenamefont {N{\'u}{\~n}ez-Regueiro}, \citenamefont
  {Tholence},\ and\ \citenamefont {Dumas}}]{fernandes2005transport}%
  \BibitemOpen
  \bibfield  {author} {\bibinfo {author} {\bibfnamefont {J.}~\bibnamefont
  {Fernandes}}, \bibinfo {author} {\bibfnamefont {R.}~\bibnamefont
  {Guimar{\~a}es}}, \bibinfo {author} {\bibfnamefont {M.}~\bibnamefont
  {Continentino}}, \bibinfo {author} {\bibfnamefont {E.}~\bibnamefont
  {Ziemath}}, \bibinfo {author} {\bibfnamefont {L.}~\bibnamefont {Walmsley}},
  \bibinfo {author} {\bibfnamefont {M.}~\bibnamefont {Monteverde}}, \bibinfo
  {author} {\bibfnamefont {M.}~\bibnamefont {N{\'u}{\~n}ez-Regueiro}}, \bibinfo
  {author} {\bibfnamefont {J.-L.}\ \bibnamefont {Tholence}},\ and\ \bibinfo
  {author} {\bibfnamefont {J.}~\bibnamefont {Dumas}},\ }\bibfield  {title}
  {\bibinfo {title} {Transport properties of the transverse charge-density-wave
  system \chem{Fe_3O_2BO_3}},\ }\href
  {https://doi.org/10.1103/PhysRevB.72.075133} {\bibfield  {journal} {\bibinfo
  {journal} {Physical Review B}\ }\textbf {\bibinfo {volume} {72}},\ \bibinfo
  {pages} {075133} (\bibinfo {year} {2005})}\BibitemShut {NoStop}%
\bibitem [{\citenamefont {Mir}\ \emph {et~al.}(2006)\citenamefont {Mir},
  \citenamefont {Janczak},\ and\ \citenamefont {Mascarenhas}}]{mir2006x}%
  \BibitemOpen
  \bibfield  {author} {\bibinfo {author} {\bibfnamefont {M.}~\bibnamefont
  {Mir}}, \bibinfo {author} {\bibfnamefont {J.}~\bibnamefont {Janczak}},\ and\
  \bibinfo {author} {\bibfnamefont {Y.}~\bibnamefont {Mascarenhas}},\
  }\bibfield  {title} {\bibinfo {title} {X-ray diffraction single-crystal
  structure characterization of iron ludwigite from room temperature to 15
  \chem{K}},\ }\href {https://doi.org/10.1107/S0021889805036344} {\bibfield
  {journal} {\bibinfo  {journal} {Journal of applied crystallography}\ }\textbf
  {\bibinfo {volume} {39}},\ \bibinfo {pages} {42} (\bibinfo {year}
  {2006})}\BibitemShut {NoStop}%
\bibitem [{\citenamefont {Ivanova}\ \emph {et~al.}(2007)\citenamefont
  {Ivanova}, \citenamefont {Vasil’ev}, \citenamefont {Velikanov},
  \citenamefont {Kazak}, \citenamefont {Ovchinnikov}, \citenamefont
  {Petrakovski{\u{\i}}},\ and\ \citenamefont {Rudenko}}]{ivanova2007magnetic}%
  \BibitemOpen
  \bibfield  {author} {\bibinfo {author} {\bibfnamefont {N.}~\bibnamefont
  {Ivanova}}, \bibinfo {author} {\bibfnamefont {A.}~\bibnamefont {Vasil’ev}},
  \bibinfo {author} {\bibfnamefont {D.}~\bibnamefont {Velikanov}}, \bibinfo
  {author} {\bibfnamefont {N.}~\bibnamefont {Kazak}}, \bibinfo {author}
  {\bibfnamefont {S.}~\bibnamefont {Ovchinnikov}}, \bibinfo {author}
  {\bibfnamefont {G.}~\bibnamefont {Petrakovski{\u{\i}}}},\ and\ \bibinfo
  {author} {\bibfnamefont {V.}~\bibnamefont {Rudenko}},\ }\bibfield  {title}
  {\bibinfo {title} {Magnetic and electrical properties of cobalt oxyborate
  \chem{Co_3BO_5}},\ }\href {https://doi.org/10.1134/S1063783407040087}
  {\bibfield  {journal} {\bibinfo  {journal} {Physics of the Solid State}\
  }\textbf {\bibinfo {volume} {49}},\ \bibinfo {pages} {651} (\bibinfo {year}
  {2007})}\BibitemShut {NoStop}%
\bibitem [{\citenamefont {de~Freitas}\ \emph {et~al.}(2008)\citenamefont
  {de~Freitas}, \citenamefont {Continentino}, \citenamefont {Guimar{\~a}es},
  \citenamefont {Fernandes}, \citenamefont {Ellena},\ and\ \citenamefont
  {Ghivelder}}]{freitas2008structure}%
  \BibitemOpen
  \bibfield  {author} {\bibinfo {author} {\bibfnamefont {D.~C.}\ \bibnamefont
  {de~Freitas}}, \bibinfo {author} {\bibfnamefont {M.~A.}\ \bibnamefont
  {Continentino}}, \bibinfo {author} {\bibfnamefont {R.~B.}\ \bibnamefont
  {Guimar{\~a}es}}, \bibinfo {author} {\bibfnamefont {J.~C.}\ \bibnamefont
  {Fernandes}}, \bibinfo {author} {\bibfnamefont {J.}~\bibnamefont {Ellena}},\
  and\ \bibinfo {author} {\bibfnamefont {L.}~\bibnamefont {Ghivelder}},\
  }\bibfield  {title} {\bibinfo {title} {Structure and magnetism of
  homometallic ludwigites: \chem{Co_2OBO_3} versus \chem{Fe_2OBO_3}},\ }\href
  {https://doi.org/10.1103/PhysRevB.77.184422} {\bibfield  {journal} {\bibinfo
  {journal} {Physical Review B}\ }\textbf {\bibinfo {volume} {77}},\ \bibinfo
  {pages} {184422} (\bibinfo {year} {2008})}\BibitemShut {NoStop}%
\bibitem [{\citenamefont {Bordet}\ and\ \citenamefont
  {Suard}(2009)}]{bordet2009magnetic}%
  \BibitemOpen
  \bibfield  {author} {\bibinfo {author} {\bibfnamefont {P.}~\bibnamefont
  {Bordet}}\ and\ \bibinfo {author} {\bibfnamefont {E.}~\bibnamefont {Suard}},\
  }\bibfield  {title} {\bibinfo {title} {Magnetic structure and charge ordering
  in \chem{Fe_3BO_5}: A single-crystal \chem{X}-ray and neutron powder
  diffraction study},\ }\href {https://doi.org/10.1103/PhysRevB.79.144408}
  {\bibfield  {journal} {\bibinfo  {journal} {Physical Review B}\ }\textbf
  {\bibinfo {volume} {79}},\ \bibinfo {pages} {144408} (\bibinfo {year}
  {2009})}\BibitemShut {NoStop}%
\bibitem [{\citenamefont {Freitas}\ \emph {et~al.}(2009)\citenamefont
  {Freitas}, \citenamefont {Continentino}, \citenamefont {Guimaraes},
  \citenamefont {Fernandes}, \citenamefont {Oliveira}, \citenamefont
  {Santelli}, \citenamefont {Ellena}, \citenamefont {Eslava},\ and\
  \citenamefont {Ghivelder}}]{freitas2009partial}%
  \BibitemOpen
  \bibfield  {author} {\bibinfo {author} {\bibfnamefont {D.}~\bibnamefont
  {Freitas}}, \bibinfo {author} {\bibfnamefont {M.~A.}\ \bibnamefont
  {Continentino}}, \bibinfo {author} {\bibfnamefont {R.~B.}\ \bibnamefont
  {Guimaraes}}, \bibinfo {author} {\bibfnamefont {J.~C.}\ \bibnamefont
  {Fernandes}}, \bibinfo {author} {\bibfnamefont {E.~P.}\ \bibnamefont
  {Oliveira}}, \bibinfo {author} {\bibfnamefont {R.}~\bibnamefont {Santelli}},
  \bibinfo {author} {\bibfnamefont {J.}~\bibnamefont {Ellena}}, \bibinfo
  {author} {\bibfnamefont {G.~G.}\ \bibnamefont {Eslava}},\ and\ \bibinfo
  {author} {\bibfnamefont {L.}~\bibnamefont {Ghivelder}},\ }\bibfield  {title}
  {\bibinfo {title} {Partial magnetic ordering and crystal structure of the
  ludwigites \chem{Co_2FeO_2BO_3} and \chem{Ni_2FeO_2BO_3}},\ }\href
  {https://doi.org/10.1103/PhysRevB.79.134437} {\bibfield  {journal} {\bibinfo
  {journal} {Physical Review B}\ }\textbf {\bibinfo {volume} {79}},\ \bibinfo
  {pages} {134437} (\bibinfo {year} {2009})}\BibitemShut {NoStop}%
\bibitem [{\citenamefont {Kazak}\ \emph
  {et~al.}(2009{\natexlab{a}})\citenamefont {Kazak}, \citenamefont {Ivanova},
  \citenamefont {Rudenko}, \citenamefont {Ovchinnikov}, \citenamefont
  {Vasil’ev},\ and\ \citenamefont {Knyazev}}]{kazak2009conductivity}%
  \BibitemOpen
  \bibfield  {author} {\bibinfo {author} {\bibfnamefont {N.}~\bibnamefont
  {Kazak}}, \bibinfo {author} {\bibfnamefont {N.}~\bibnamefont {Ivanova}},
  \bibinfo {author} {\bibfnamefont {V.}~\bibnamefont {Rudenko}}, \bibinfo
  {author} {\bibfnamefont {S.}~\bibnamefont {Ovchinnikov}}, \bibinfo {author}
  {\bibfnamefont {A.}~\bibnamefont {Vasil’ev}},\ and\ \bibinfo {author}
  {\bibfnamefont {Y.~V.}\ \bibnamefont {Knyazev}},\ }\bibfield  {title}
  {\bibinfo {title} {Conductivity study of \chem{Co_3O_2BO_3} and
  \chem{Co_{3-x}Fe_xO_2BO_3} oxyborates},\ }in\ \href
  {https://doi.org/10.4028/www.scientific.net/SSP.152-153.104} {\emph {\bibinfo
  {booktitle} {Solid State Phenomena}}},\ Vol.\ \bibinfo {volume} {152}\
  (\bibinfo {organization} {Trans Tech Publ},\ \bibinfo {year} {2009})\ pp.\
  \bibinfo {pages} {104--107}\BibitemShut {NoStop}%
\bibitem [{\citenamefont {Kazak}\ \emph
  {et~al.}(2009{\natexlab{b}})\citenamefont {Kazak}, \citenamefont {Ivanova},
  \citenamefont {Rudenko}, \citenamefont {Vasil’ev}, \citenamefont
  {Velikanov},\ and\ \citenamefont {Ovchinnikov}}]{kazak2009low}%
  \BibitemOpen
  \bibfield  {author} {\bibinfo {author} {\bibfnamefont {N.}~\bibnamefont
  {Kazak}}, \bibinfo {author} {\bibfnamefont {N.}~\bibnamefont {Ivanova}},
  \bibinfo {author} {\bibfnamefont {V.}~\bibnamefont {Rudenko}}, \bibinfo
  {author} {\bibfnamefont {A.}~\bibnamefont {Vasil’ev}}, \bibinfo {author}
  {\bibfnamefont {D.}~\bibnamefont {Velikanov}},\ and\ \bibinfo {author}
  {\bibfnamefont {S.}~\bibnamefont {Ovchinnikov}},\ }\bibfield  {title}
  {\bibinfo {title} {Low-field magnetization of ludwigites \chem{Co_3O_2BO_3}
  and \chem{Co_{3-x}Fe_xO_2BO_3} ($x\approx0.14$)},\ }\href
  {https://doi.org/10.1134/S1063783409050138} {\bibfield  {journal} {\bibinfo
  {journal} {Physics of the Solid State}\ }\textbf {\bibinfo {volume} {51}},\
  \bibinfo {pages} {966} (\bibinfo {year} {2009}{\natexlab{b}})}\BibitemShut
  {NoStop}%
\bibitem [{\citenamefont {Freitas}\ \emph {et~al.}(2010)\citenamefont
  {Freitas}, \citenamefont {Guimaraes}, \citenamefont {Sanchez}, \citenamefont
  {Fernandes}, \citenamefont {Continentino}, \citenamefont {Ellena},
  \citenamefont {Kitada}, \citenamefont {Kageyama}, \citenamefont {Matsuo},
  \citenamefont {Kindo} \emph {et~al.}}]{freitas2010structural}%
  \BibitemOpen
  \bibfield  {author} {\bibinfo {author} {\bibfnamefont {D.}~\bibnamefont
  {Freitas}}, \bibinfo {author} {\bibfnamefont {R.~B.}\ \bibnamefont
  {Guimaraes}}, \bibinfo {author} {\bibfnamefont {D.~R.}\ \bibnamefont
  {Sanchez}}, \bibinfo {author} {\bibfnamefont {J.~C.}\ \bibnamefont
  {Fernandes}}, \bibinfo {author} {\bibfnamefont {M.~A.}\ \bibnamefont
  {Continentino}}, \bibinfo {author} {\bibfnamefont {J.}~\bibnamefont
  {Ellena}}, \bibinfo {author} {\bibfnamefont {A.}~\bibnamefont {Kitada}},
  \bibinfo {author} {\bibfnamefont {H.}~\bibnamefont {Kageyama}}, \bibinfo
  {author} {\bibfnamefont {A.}~\bibnamefont {Matsuo}}, \bibinfo {author}
  {\bibfnamefont {K.}~\bibnamefont {Kindo}}, \emph {et~al.},\ }\bibfield
  {title} {\bibinfo {title} {Structural and magnetic properties of the
  oxyborate \chem{Co_5Ti(O_2BO_3)_2}},\ }\href
  {https://doi.org/10.1103/PhysRevB.81.024432} {\bibfield  {journal} {\bibinfo
  {journal} {Physical Review B}\ }\textbf {\bibinfo {volume} {81}},\ \bibinfo
  {pages} {024432} (\bibinfo {year} {2010})}\BibitemShut {NoStop}%
\bibitem [{\citenamefont {Kazak}\ \emph {et~al.}(2011)\citenamefont {Kazak},
  \citenamefont {Ivanova}, \citenamefont {Bayukov}, \citenamefont
  {Ovchinnikov}, \citenamefont {Vasiliev}, \citenamefont {Rudenko},
  \citenamefont {Bartolom{\'e}}, \citenamefont {Arauzo},\ and\ \citenamefont
  {Knyazev}}]{kazak2011superexchange}%
  \BibitemOpen
  \bibfield  {author} {\bibinfo {author} {\bibfnamefont {N.}~\bibnamefont
  {Kazak}}, \bibinfo {author} {\bibfnamefont {N.}~\bibnamefont {Ivanova}},
  \bibinfo {author} {\bibfnamefont {O.}~\bibnamefont {Bayukov}}, \bibinfo
  {author} {\bibfnamefont {S.}~\bibnamefont {Ovchinnikov}}, \bibinfo {author}
  {\bibfnamefont {A.}~\bibnamefont {Vasiliev}}, \bibinfo {author}
  {\bibfnamefont {V.}~\bibnamefont {Rudenko}}, \bibinfo {author} {\bibfnamefont
  {J.}~\bibnamefont {Bartolom{\'e}}}, \bibinfo {author} {\bibfnamefont
  {A.}~\bibnamefont {Arauzo}},\ and\ \bibinfo {author} {\bibfnamefont {Y.~V.}\
  \bibnamefont {Knyazev}},\ }\bibfield  {title} {\bibinfo {title} {The
  superexchange interactions in mixed \chem{Co}--\chem{Fe} ludwigite},\ }\href
  {https://doi.org/10.1016/j.jmmm.2010.09.057} {\bibfield  {journal} {\bibinfo
  {journal} {Journal of Magnetism and Magnetic Materials}\ }\textbf {\bibinfo
  {volume} {323}},\ \bibinfo {pages} {521} (\bibinfo {year}
  {2011})}\BibitemShut {NoStop}%
\bibitem [{\citenamefont {Bartolom{\'e}}\ \emph {et~al.}(2011)\citenamefont
  {Bartolom{\'e}}, \citenamefont {Arauzo}, \citenamefont {Kazak}, \citenamefont
  {Ivanova}, \citenamefont {Ovchinnikov}, \citenamefont {Knyazev},\ and\
  \citenamefont {Lyubutin}}]{bartolome2011uniaxial}%
  \BibitemOpen
  \bibfield  {author} {\bibinfo {author} {\bibfnamefont {J.}~\bibnamefont
  {Bartolom{\'e}}}, \bibinfo {author} {\bibfnamefont {A.}~\bibnamefont
  {Arauzo}}, \bibinfo {author} {\bibfnamefont {N.~V.}\ \bibnamefont {Kazak}},
  \bibinfo {author} {\bibfnamefont {N.~B.}\ \bibnamefont {Ivanova}}, \bibinfo
  {author} {\bibfnamefont {S.~G.}\ \bibnamefont {Ovchinnikov}}, \bibinfo
  {author} {\bibfnamefont {Y.~V.}\ \bibnamefont {Knyazev}},\ and\ \bibinfo
  {author} {\bibfnamefont {I.~S.}\ \bibnamefont {Lyubutin}},\ }\bibfield
  {title} {\bibinfo {title} {Uniaxial magnetic anisotropy in
  \chem{Co_{2.25}Fe_{0.75}O_2BO_3} compared to \chem{Co_3O_2BO_3} and
  \chem{Fe_3O_2BO} ludwigites},\ }\href
  {https://doi.org/10.1103/PhysRevB.83.144426} {\bibfield  {journal} {\bibinfo
  {journal} {Physical Review B}\ }\textbf {\bibinfo {volume} {83}},\ \bibinfo
  {pages} {144426} (\bibinfo {year} {2011})}\BibitemShut {NoStop}%
\bibitem [{\citenamefont {Ivanova}\ \emph {et~al.}(2011)\citenamefont
  {Ivanova}, \citenamefont {Kazak}, \citenamefont {Knyazev}, \citenamefont
  {Velikanov}, \citenamefont {Bezmaternykh}, \citenamefont {Ovchinnikov},
  \citenamefont {Vasiliev}, \citenamefont {Platunov}, \citenamefont
  {Bartolom{\'e}},\ and\ \citenamefont {Patrin}}]{ivanova2011crystal}%
  \BibitemOpen
  \bibfield  {author} {\bibinfo {author} {\bibfnamefont {N.}~\bibnamefont
  {Ivanova}}, \bibinfo {author} {\bibfnamefont {N.}~\bibnamefont {Kazak}},
  \bibinfo {author} {\bibfnamefont {Y.~V.}\ \bibnamefont {Knyazev}}, \bibinfo
  {author} {\bibfnamefont {D.}~\bibnamefont {Velikanov}}, \bibinfo {author}
  {\bibfnamefont {L.}~\bibnamefont {Bezmaternykh}}, \bibinfo {author}
  {\bibfnamefont {S.}~\bibnamefont {Ovchinnikov}}, \bibinfo {author}
  {\bibfnamefont {A.}~\bibnamefont {Vasiliev}}, \bibinfo {author}
  {\bibfnamefont {M.}~\bibnamefont {Platunov}}, \bibinfo {author}
  {\bibfnamefont {J.}~\bibnamefont {Bartolom{\'e}}},\ and\ \bibinfo {author}
  {\bibfnamefont {G.}~\bibnamefont {Patrin}},\ }\bibfield  {title} {\bibinfo
  {title} {Crystal structure and magnetic anisotropy of ludwigite
  \chem{Co_2FeO_2BO_3}},\ }\href {https://doi.org/10.1134/S1063776111140172}
  {\bibfield  {journal} {\bibinfo  {journal} {Journal of Experimental and
  Theoretical Physics}\ }\textbf {\bibinfo {volume} {113}},\ \bibinfo {pages}
  {1015} (\bibinfo {year} {2011})}\BibitemShut {NoStop}%
\bibitem [{\citenamefont {Knyazev}\ \emph {et~al.}(2012)\citenamefont
  {Knyazev}, \citenamefont {Ivanova}, \citenamefont {Kazak}, \citenamefont
  {Platunov}, \citenamefont {Bezmaternykh}, \citenamefont {Velikanov},
  \citenamefont {Vasiliev}, \citenamefont {Ovchinnikov},\ and\ \citenamefont
  {Yurkin}}]{knyazev2012crystal}%
  \BibitemOpen
  \bibfield  {author} {\bibinfo {author} {\bibfnamefont {Y.~V.}\ \bibnamefont
  {Knyazev}}, \bibinfo {author} {\bibfnamefont {N.}~\bibnamefont {Ivanova}},
  \bibinfo {author} {\bibfnamefont {N.}~\bibnamefont {Kazak}}, \bibinfo
  {author} {\bibfnamefont {M.}~\bibnamefont {Platunov}}, \bibinfo {author}
  {\bibfnamefont {L.}~\bibnamefont {Bezmaternykh}}, \bibinfo {author}
  {\bibfnamefont {D.}~\bibnamefont {Velikanov}}, \bibinfo {author}
  {\bibfnamefont {A.}~\bibnamefont {Vasiliev}}, \bibinfo {author}
  {\bibfnamefont {S.}~\bibnamefont {Ovchinnikov}},\ and\ \bibinfo {author}
  {\bibfnamefont {G.~Y.}\ \bibnamefont {Yurkin}},\ }\bibfield  {title}
  {\bibinfo {title} {Crystal structure and magnetic properties of \chem{Mn}
  substituted ludwigite \chem{Co_3O_2BO_3}},\ }\href
  {https://doi.org/https://doi.org/10.1016/j.jmmm.2011.07.044} {\bibfield
  {journal} {\bibinfo  {journal} {Journal of Magnetism and Magnetic Materials}\
  }\textbf {\bibinfo {volume} {324}},\ \bibinfo {pages} {923} (\bibinfo {year}
  {2012})}\BibitemShut {NoStop}%
\bibitem [{\citenamefont {Ivanova}\ \emph
  {et~al.}(2012{\natexlab{a}})\citenamefont {Ivanova}, \citenamefont
  {Platunov}, \citenamefont {Knyazev}, \citenamefont {Kazak}, \citenamefont
  {Bezmaternykh}, \citenamefont {Vasiliev}, \citenamefont {Ovchinnikov},\ and\
  \citenamefont {Nizhankovskii}}]{ivanova2012effect}%
  \BibitemOpen
  \bibfield  {author} {\bibinfo {author} {\bibfnamefont {N.}~\bibnamefont
  {Ivanova}}, \bibinfo {author} {\bibfnamefont {M.}~\bibnamefont {Platunov}},
  \bibinfo {author} {\bibfnamefont {Y.~V.}\ \bibnamefont {Knyazev}}, \bibinfo
  {author} {\bibfnamefont {N.}~\bibnamefont {Kazak}}, \bibinfo {author}
  {\bibfnamefont {L.}~\bibnamefont {Bezmaternykh}}, \bibinfo {author}
  {\bibfnamefont {A.}~\bibnamefont {Vasiliev}}, \bibinfo {author}
  {\bibfnamefont {S.}~\bibnamefont {Ovchinnikov}},\ and\ \bibinfo {author}
  {\bibfnamefont {V.}~\bibnamefont {Nizhankovskii}},\ }\bibfield  {title}
  {\bibinfo {title} {Effect of the diamagnetic dilution on the magnetic
  ordering and electrical conductivity in the \chem{Co_3O_2BO_3}:\chem{Ga}
  ludwigite},\ }\href {https://doi.org/10.1134/S1063783412110133} {\bibfield
  {journal} {\bibinfo  {journal} {Physics of the Solid State}\ }\textbf
  {\bibinfo {volume} {54}},\ \bibinfo {pages} {2212} (\bibinfo {year}
  {2012}{\natexlab{a}})}\BibitemShut {NoStop}%
\bibitem [{\citenamefont {Ivanova}\ \emph
  {et~al.}(2012{\natexlab{b}})\citenamefont {Ivanova}, \citenamefont
  {Platunov}, \citenamefont {Knyazev}, \citenamefont {Kazak}, \citenamefont
  {Bezmaternykh}, \citenamefont {Eremin},\ and\ \citenamefont
  {Vasiliev}}]{ivanova2012spin}%
  \BibitemOpen
  \bibfield  {author} {\bibinfo {author} {\bibfnamefont {N.}~\bibnamefont
  {Ivanova}}, \bibinfo {author} {\bibfnamefont {M.}~\bibnamefont {Platunov}},
  \bibinfo {author} {\bibfnamefont {Y.~V.}\ \bibnamefont {Knyazev}}, \bibinfo
  {author} {\bibfnamefont {N.}~\bibnamefont {Kazak}}, \bibinfo {author}
  {\bibfnamefont {L.}~\bibnamefont {Bezmaternykh}}, \bibinfo {author}
  {\bibfnamefont {E.}~\bibnamefont {Eremin}},\ and\ \bibinfo {author}
  {\bibfnamefont {A.}~\bibnamefont {Vasiliev}},\ }\bibfield  {title} {\bibinfo
  {title} {Spin-glass magnetic ordering in \chem{CoMgGaO_2BO_3} ludwigite},\
  }\href {https://doi.org/10.1063/1.3679627} {\bibfield  {journal} {\bibinfo
  {journal} {Low Temperature Physics}\ }\textbf {\bibinfo {volume} {38}},\
  \bibinfo {pages} {172} (\bibinfo {year} {2012}{\natexlab{b}})}\BibitemShut
  {NoStop}%
\bibitem [{\citenamefont {Ivanova}\ \emph {et~al.}(2013)\citenamefont
  {Ivanova}, \citenamefont {Kazak}, \citenamefont {Knyazev}, \citenamefont
  {Velikanov}, \citenamefont {Vasiliev}, \citenamefont {Bezmaternykh},\ and\
  \citenamefont {Platunov}}]{ivanova2013structure}%
  \BibitemOpen
  \bibfield  {author} {\bibinfo {author} {\bibfnamefont {N.}~\bibnamefont
  {Ivanova}}, \bibinfo {author} {\bibfnamefont {N.}~\bibnamefont {Kazak}},
  \bibinfo {author} {\bibfnamefont {Y.~V.}\ \bibnamefont {Knyazev}}, \bibinfo
  {author} {\bibfnamefont {D.}~\bibnamefont {Velikanov}}, \bibinfo {author}
  {\bibfnamefont {A.}~\bibnamefont {Vasiliev}}, \bibinfo {author}
  {\bibfnamefont {L.}~\bibnamefont {Bezmaternykh}},\ and\ \bibinfo {author}
  {\bibfnamefont {M.}~\bibnamefont {Platunov}},\ }\bibfield  {title} {\bibinfo
  {title} {Structure and magnetism of copper-substituted cobalt ludwigite
  \chem{Co_3O_2BO_3}},\ }\href {https://doi.org/10.1063/1.4818633} {\bibfield
  {journal} {\bibinfo  {journal} {Low Temperature Physics}\ }\textbf {\bibinfo
  {volume} {39}},\ \bibinfo {pages} {709} (\bibinfo {year} {2013})}\BibitemShut
  {NoStop}%
\bibitem [{\citenamefont {Medrano}\ \emph {et~al.}(2015)\citenamefont
  {Medrano}, \citenamefont {Freitas}, \citenamefont {Sanchez}, \citenamefont
  {Pinheiro}, \citenamefont {Eslava}, \citenamefont {Ghivelder},\ and\
  \citenamefont {Continentino}}]{medrano2015nonmagnetic}%
  \BibitemOpen
  \bibfield  {author} {\bibinfo {author} {\bibfnamefont {C.~P.~C.}\
  \bibnamefont {Medrano}}, \bibinfo {author} {\bibfnamefont {D.~C.}\
  \bibnamefont {Freitas}}, \bibinfo {author} {\bibfnamefont {D.~R.}\
  \bibnamefont {Sanchez}}, \bibinfo {author} {\bibfnamefont {C.~B.}\
  \bibnamefont {Pinheiro}}, \bibinfo {author} {\bibfnamefont {G.~G.}\
  \bibnamefont {Eslava}}, \bibinfo {author} {\bibfnamefont {L.}~\bibnamefont
  {Ghivelder}},\ and\ \bibinfo {author} {\bibfnamefont {M.~A.}\ \bibnamefont
  {Continentino}},\ }\bibfield  {title} {\bibinfo {title} {Nonmagnetic ions
  enhance magnetic order in the ludwigite \chem{Co_5Sn(O_2BO_3)_2}},\ }\href
  {https://doi.org/10.1103/PhysRevB.91.054402} {\bibfield  {journal} {\bibinfo
  {journal} {Phys. Rev. B}\ }\textbf {\bibinfo {volume} {91}},\ \bibinfo
  {pages} {054402} (\bibinfo {year} {2015})}\BibitemShut {NoStop}%
\bibitem [{\citenamefont {dos Santos}\ \emph {et~al.}(2016)\citenamefont {dos
  Santos}, \citenamefont {Freitas}, \citenamefont {Fier}, \citenamefont
  {Fernandes}, \citenamefont {Continentino}, \citenamefont {De~Oliveira},\ and\
  \citenamefont {Walmsley}}]{dos2016current}%
  \BibitemOpen
  \bibfield  {author} {\bibinfo {author} {\bibfnamefont {E.}~\bibnamefont {dos
  Santos}}, \bibinfo {author} {\bibfnamefont {D.}~\bibnamefont {Freitas}},
  \bibinfo {author} {\bibfnamefont {I.}~\bibnamefont {Fier}}, \bibinfo {author}
  {\bibfnamefont {J.}~\bibnamefont {Fernandes}}, \bibinfo {author}
  {\bibfnamefont {M.}~\bibnamefont {Continentino}}, \bibinfo {author}
  {\bibfnamefont {A.}~\bibnamefont {De~Oliveira}},\ and\ \bibinfo {author}
  {\bibfnamefont {L.}~\bibnamefont {Walmsley}},\ }\bibfield  {title} {\bibinfo
  {title} {Current controlled negative differential resistance behavior in
  \chem{Co_2FeO_2BO_3} and \chem{Fe_3O_2BO_3} single crystals},\ }\href
  {https://doi.org/10.1016/j.jpcs.2015.11.015} {\bibfield  {journal} {\bibinfo
  {journal} {Journal of Physics and Chemistry of Solids}\ }\textbf {\bibinfo
  {volume} {90}},\ \bibinfo {pages} {65} (\bibinfo {year} {2016})}\BibitemShut
  {NoStop}%
\bibitem [{\citenamefont {Freitas}\ \emph {et~al.}(2016)\citenamefont
  {Freitas}, \citenamefont {Medrano}, \citenamefont {Sanchez}, \citenamefont
  {Regueiro}, \citenamefont {Rodr{\'i}guez-Velamaz{\'a}n},\ and\ \citenamefont
  {Continentino}}]{freitas2016magnetism}%
  \BibitemOpen
  \bibfield  {author} {\bibinfo {author} {\bibfnamefont {D.~C.}\ \bibnamefont
  {Freitas}}, \bibinfo {author} {\bibfnamefont {C.~P.~C.}\ \bibnamefont
  {Medrano}}, \bibinfo {author} {\bibfnamefont {D.~R.}\ \bibnamefont
  {Sanchez}}, \bibinfo {author} {\bibfnamefont {M.~N.}\ \bibnamefont
  {Regueiro}}, \bibinfo {author} {\bibfnamefont {J.~A.}\ \bibnamefont
  {Rodr{\'i}guez-Velamaz{\'a}n}},\ and\ \bibinfo {author} {\bibfnamefont
  {M.~A.}\ \bibnamefont {Continentino}},\ }\bibfield  {title} {\bibinfo {title}
  {Magnetism and charge order in the ladder compound \chem{Co_3O_2BO_3}},\
  }\href {https://doi.org/10.1103/PhysRevB.94.174409} {\bibfield  {journal}
  {\bibinfo  {journal} {Physical Review B}\ }\textbf {\bibinfo {volume} {94}},\
  \bibinfo {pages} {174409} (\bibinfo {year} {2016})}\BibitemShut {NoStop}%
\bibitem [{\citenamefont {Sofronova}\ and\ \citenamefont
  {Nazarenko}(2017)}]{sofronova2017ludwigites}%
  \BibitemOpen
  \bibfield  {author} {\bibinfo {author} {\bibfnamefont {S.}~\bibnamefont
  {Sofronova}}\ and\ \bibinfo {author} {\bibfnamefont {I.}~\bibnamefont
  {Nazarenko}},\ }\bibfield  {title} {\bibinfo {title} {Ludwigites: From
  natural mineral to modern solid solutions},\ }\href
  {https://doi.org/10.1002/crat.201600338} {\bibfield  {journal} {\bibinfo
  {journal} {Crystal Research and Technology}\ }\textbf {\bibinfo {volume}
  {52}},\ \bibinfo {pages} {1600338} (\bibinfo {year} {2017})}\BibitemShut
  {NoStop}%
\bibitem [{\citenamefont {Medrano}\ \emph {et~al.}(2017)\citenamefont
  {Medrano}, \citenamefont {Freitas}, \citenamefont {Passamani}, \citenamefont
  {Pinheiro}, \citenamefont {Baggio-Saitovitch}, \citenamefont {Continentino},\
  and\ \citenamefont {Sanchez}}]{medrano2017}%
  \BibitemOpen
  \bibfield  {author} {\bibinfo {author} {\bibfnamefont {C.~P.~C.}\
  \bibnamefont {Medrano}}, \bibinfo {author} {\bibfnamefont {D.~C.}\
  \bibnamefont {Freitas}}, \bibinfo {author} {\bibfnamefont {E.~C.}\
  \bibnamefont {Passamani}}, \bibinfo {author} {\bibfnamefont {C.~B.}\
  \bibnamefont {Pinheiro}}, \bibinfo {author} {\bibfnamefont {E.}~\bibnamefont
  {Baggio-Saitovitch}}, \bibinfo {author} {\bibfnamefont {M.~A.}\ \bibnamefont
  {Continentino}},\ and\ \bibinfo {author} {\bibfnamefont {D.~R.}\ \bibnamefont
  {Sanchez}},\ }\bibfield  {title} {\bibinfo {title} {Field-induced
  metamagnetic transitions and two-dimensional excitations in ludwigite
  \chem{Co_{4.76}Al_{1.24}(O_2BO_3)_2}},\ }\href
  {https://doi.org/10.1103/PhysRevB.95.214419} {\bibfield  {journal} {\bibinfo
  {journal} {Phys. Rev. B}\ }\textbf {\bibinfo {volume} {95}},\ \bibinfo
  {pages} {214419} (\bibinfo {year} {2017})}\BibitemShut {NoStop}%
\bibitem [{\citenamefont {Kumar}\ \emph {et~al.}(2017)\citenamefont {Kumar},
  \citenamefont {Panja}, \citenamefont {Mukkattukavil}, \citenamefont
  {Bhattacharyya}, \citenamefont {Nigam},\ and\ \citenamefont
  {Nair}}]{kumar2017reentrant}%
  \BibitemOpen
  \bibfield  {author} {\bibinfo {author} {\bibfnamefont {J.}~\bibnamefont
  {Kumar}}, \bibinfo {author} {\bibfnamefont {S.~N.}\ \bibnamefont {Panja}},
  \bibinfo {author} {\bibfnamefont {D.~J.}\ \bibnamefont {Mukkattukavil}},
  \bibinfo {author} {\bibfnamefont {A.}~\bibnamefont {Bhattacharyya}}, \bibinfo
  {author} {\bibfnamefont {A.}~\bibnamefont {Nigam}},\ and\ \bibinfo {author}
  {\bibfnamefont {S.}~\bibnamefont {Nair}},\ }\bibfield  {title} {\bibinfo
  {title} {Reentrant superspin glass state and magnetization steps in the
  oxyborate \chem{Co_2AlBO_5}},\ }\href
  {https://doi.org/10.1103/PhysRevB.95.144409} {\bibfield  {journal} {\bibinfo
  {journal} {Physical Review B}\ }\textbf {\bibinfo {volume} {95}},\ \bibinfo
  {pages} {144409} (\bibinfo {year} {2017})}\BibitemShut {NoStop}%
\bibitem [{\citenamefont {dos Santos}\ \emph {et~al.}(2017)\citenamefont {dos
  Santos}, \citenamefont {da~Silva}, \citenamefont {Fernandes}, \citenamefont
  {Ghivelder}, \citenamefont {Freitas}, \citenamefont {Continentino},\ and\
  \citenamefont {Walmsley}}]{dos2017non}%
  \BibitemOpen
  \bibfield  {author} {\bibinfo {author} {\bibfnamefont {E.}~\bibnamefont {dos
  Santos}}, \bibinfo {author} {\bibfnamefont {E.}~\bibnamefont {da~Silva}},
  \bibinfo {author} {\bibfnamefont {J.}~\bibnamefont {Fernandes}}, \bibinfo
  {author} {\bibfnamefont {L.}~\bibnamefont {Ghivelder}}, \bibinfo {author}
  {\bibfnamefont {D.}~\bibnamefont {Freitas}}, \bibinfo {author} {\bibfnamefont
  {M.}~\bibnamefont {Continentino}},\ and\ \bibinfo {author} {\bibfnamefont
  {L.}~\bibnamefont {Walmsley}},\ }\bibfield  {title} {\bibinfo {title}
  {Non-linear conduction due to depinning of charge order domains in
  \chem{Fe_3O_2BO_3}},\ }\href {https://doi.org/10.1088/1361-648X/aa6960}
  {\bibfield  {journal} {\bibinfo  {journal} {Journal of Physics: Condensed
  Matter}\ }\textbf {\bibinfo {volume} {29}},\ \bibinfo {pages} {205401}
  (\bibinfo {year} {2017})}\BibitemShut {NoStop}%
\bibitem [{\citenamefont {Galdino}\ \emph {et~al.}(2019)\citenamefont
  {Galdino}, \citenamefont {Freitas}, \citenamefont {Medrano}, \citenamefont
  {Tartaglia}, \citenamefont {Rigitano}, \citenamefont {Oliveira},
  \citenamefont {Mendon{\c{c}}a}, \citenamefont {Ghivelder}, \citenamefont
  {Continentino}, \citenamefont {Sanchez},\ and\ \citenamefont
  {Granado}}]{galdino2019magnetic}%
  \BibitemOpen
  \bibfield  {author} {\bibinfo {author} {\bibfnamefont {C.~W.}\ \bibnamefont
  {Galdino}}, \bibinfo {author} {\bibfnamefont {D.~C.}\ \bibnamefont
  {Freitas}}, \bibinfo {author} {\bibfnamefont {C.~P.~C.}\ \bibnamefont
  {Medrano}}, \bibinfo {author} {\bibfnamefont {R.}~\bibnamefont {Tartaglia}},
  \bibinfo {author} {\bibfnamefont {D.}~\bibnamefont {Rigitano}}, \bibinfo
  {author} {\bibfnamefont {J.~F.}\ \bibnamefont {Oliveira}}, \bibinfo {author}
  {\bibfnamefont {A.~A.}\ \bibnamefont {Mendon{\c{c}}a}}, \bibinfo {author}
  {\bibfnamefont {L.}~\bibnamefont {Ghivelder}}, \bibinfo {author}
  {\bibfnamefont {M.~A.}\ \bibnamefont {Continentino}}, \bibinfo {author}
  {\bibfnamefont {D.~R.}\ \bibnamefont {Sanchez}},\ and\ \bibinfo {author}
  {\bibfnamefont {E.}~\bibnamefont {Granado}},\ }\bibfield  {title} {\bibinfo
  {title} {Magnetic, electronic, structural, and thermal properties of the
  \chem{Co_3O_2BO_3} ludwigite in the paramagnetic state},\ }\href
  {https://doi.org/10.1103/PhysRevB.100.165138} {\bibfield  {journal} {\bibinfo
   {journal} {Physical Review B}\ }\textbf {\bibinfo {volume} {100}},\ \bibinfo
  {pages} {165138} (\bibinfo {year} {2019})}\BibitemShut {NoStop}%
\bibitem [{\citenamefont {Knyazev}\ \emph {et~al.}(2019)\citenamefont
  {Knyazev}, \citenamefont {Kazak}, \citenamefont {Nazarenko}, \citenamefont
  {Sofronova}, \citenamefont {Rostovtsev}, \citenamefont {Bartolom{\'e}},
  \citenamefont {Arauzo},\ and\ \citenamefont
  {Ovchinnikov}}]{knyazev2019effect}%
  \BibitemOpen
  \bibfield  {author} {\bibinfo {author} {\bibfnamefont {Y.~V.}\ \bibnamefont
  {Knyazev}}, \bibinfo {author} {\bibfnamefont {N.}~\bibnamefont {Kazak}},
  \bibinfo {author} {\bibfnamefont {I.}~\bibnamefont {Nazarenko}}, \bibinfo
  {author} {\bibfnamefont {S.}~\bibnamefont {Sofronova}}, \bibinfo {author}
  {\bibfnamefont {N.}~\bibnamefont {Rostovtsev}}, \bibinfo {author}
  {\bibfnamefont {J.}~\bibnamefont {Bartolom{\'e}}}, \bibinfo {author}
  {\bibfnamefont {A.}~\bibnamefont {Arauzo}},\ and\ \bibinfo {author}
  {\bibfnamefont {S.}~\bibnamefont {Ovchinnikov}},\ }\bibfield  {title}
  {\bibinfo {title} {Effect of magnetic frustrations on magnetism of the
  \chem{Fe_3BO_5} and \chem{Co_3BO_5} ludwigites},\ }\href
  {https://doi.org/https://doi.org/10.1016/j.jmmm.2018.10.126} {\bibfield
  {journal} {\bibinfo  {journal} {Journal of Magnetism and Magnetic Materials}\
  }\textbf {\bibinfo {volume} {474}},\ \bibinfo {pages} {493} (\bibinfo {year}
  {2019})}\BibitemShut {NoStop}%
\bibitem [{\citenamefont {Mariano}\ \emph {et~al.}(2020)\citenamefont
  {Mariano}, \citenamefont {Heringer}, \citenamefont {Freitas}, \citenamefont
  {Baggio-Saitovitch}, \citenamefont {Continentino}, \citenamefont
  {Passamani},\ and\ \citenamefont {Sanchez}}]{mariano2020}%
  \BibitemOpen
  \bibfield  {author} {\bibinfo {author} {\bibfnamefont {D.~L.}\ \bibnamefont
  {Mariano}}, \bibinfo {author} {\bibfnamefont {M.~A.~V.}\ \bibnamefont
  {Heringer}}, \bibinfo {author} {\bibfnamefont {D.~C.}\ \bibnamefont
  {Freitas}}, \bibinfo {author} {\bibfnamefont {E.}~\bibnamefont
  {Baggio-Saitovitch}}, \bibinfo {author} {\bibfnamefont {M.~A.}\ \bibnamefont
  {Continentino}}, \bibinfo {author} {\bibfnamefont {E.~C.}\ \bibnamefont
  {Passamani}},\ and\ \bibinfo {author} {\bibfnamefont {D.~R.}\ \bibnamefont
  {Sanchez}},\ }\bibfield  {title} {\bibinfo {title} {Dimensional crossover in
  \chem{Cr}-doped \chem{Co_3BO_5}},\ }\href
  {https://doi.org/10.1103/PhysRevB.102.064424} {\bibfield  {journal} {\bibinfo
   {journal} {Phys. Rev. B}\ }\textbf {\bibinfo {volume} {102}},\ \bibinfo
  {pages} {064424} (\bibinfo {year} {2020})}\BibitemShut {NoStop}%
\bibitem [{\citenamefont {Heringer}\ \emph {et~al.}(2020)\citenamefont
  {Heringer}, \citenamefont {Mariano}, \citenamefont {Freitas}, \citenamefont
  {Baggio-Saitovitch}, \citenamefont {Continentino},\ and\ \citenamefont
  {Sanchez}}]{heringer2020spin}%
  \BibitemOpen
  \bibfield  {author} {\bibinfo {author} {\bibfnamefont {M.}~\bibnamefont
  {Heringer}}, \bibinfo {author} {\bibfnamefont {D.}~\bibnamefont {Mariano}},
  \bibinfo {author} {\bibfnamefont {D.}~\bibnamefont {Freitas}}, \bibinfo
  {author} {\bibfnamefont {E.}~\bibnamefont {Baggio-Saitovitch}}, \bibinfo
  {author} {\bibfnamefont {M.}~\bibnamefont {Continentino}},\ and\ \bibinfo
  {author} {\bibfnamefont {D.}~\bibnamefont {Sanchez}},\ }\bibfield  {title}
  {\bibinfo {title} {Spin-glass behavior in \chem{Co_3Mn_3(O_2BO_3)_2}
  ludwigite with weak disorder},\ }\href
  {https://doi.org/10.1103/PhysRevMaterials.4.064412} {\bibfield  {journal}
  {\bibinfo  {journal} {Physical Review Materials}\ }\textbf {\bibinfo {volume}
  {4}},\ \bibinfo {pages} {064412} (\bibinfo {year} {2020})}\BibitemShut
  {NoStop}%
\bibitem [{\citenamefont {Medrano}\ \emph {et~al.}(2021)\citenamefont
  {Medrano}, \citenamefont {Sadrollahi}, \citenamefont {Da~Fonseca},
  \citenamefont {Passamani}, \citenamefont {Freitas}, \citenamefont
  {Continentino}, \citenamefont {Sanchez}, \citenamefont {Litterst},\ and\
  \citenamefont {Baggio-Saitovitch}}]{medrano2021magnetic}%
  \BibitemOpen
  \bibfield  {author} {\bibinfo {author} {\bibfnamefont {C.}~\bibnamefont
  {Medrano}}, \bibinfo {author} {\bibfnamefont {E.}~\bibnamefont {Sadrollahi}},
  \bibinfo {author} {\bibfnamefont {R.}~\bibnamefont {Da~Fonseca}}, \bibinfo
  {author} {\bibfnamefont {E.}~\bibnamefont {Passamani}}, \bibinfo {author}
  {\bibfnamefont {D.}~\bibnamefont {Freitas}}, \bibinfo {author} {\bibfnamefont
  {M.}~\bibnamefont {Continentino}}, \bibinfo {author} {\bibfnamefont
  {D.}~\bibnamefont {Sanchez}}, \bibinfo {author} {\bibfnamefont
  {F.}~\bibnamefont {Litterst}},\ and\ \bibinfo {author} {\bibfnamefont
  {E.}~\bibnamefont {Baggio-Saitovitch}},\ }\bibfield  {title} {\bibinfo
  {title} {Magnetic properties of \chem{Ni_5Sn(O_2BO_3)_2} ludwigite},\ }\href
  {https://doi.org/10.1103/PhysRevB.103.064430} {\bibfield  {journal} {\bibinfo
   {journal} {Physical Review B}\ }\textbf {\bibinfo {volume} {103}},\ \bibinfo
  {pages} {064430} (\bibinfo {year} {2021})}\BibitemShut {NoStop}%
\bibitem [{\citenamefont {Galdino}\ \emph {et~al.}(2021)\citenamefont
  {Galdino}, \citenamefont {Freitas}, \citenamefont {Medrano}, \citenamefont
  {Sanchez}, \citenamefont {Tartaglia}, \citenamefont {Rabello}, \citenamefont
  {Mendon\ifmmode~\mbox{\c{c}}\else \c{c}\fi{}a}, \citenamefont {Ghivelder},
  \citenamefont {Continentino}, \citenamefont {Zapata}, \citenamefont
  {Pinheiro}, \citenamefont {Azevedo}, \citenamefont
  {Rodr\'{\i}guez-Velamaz\'an}, \citenamefont {Garbarino}, \citenamefont
  {N\'u\~nez Regueiro},\ and\ \citenamefont {Granado}}]{galdino2021}%
  \BibitemOpen
  \bibfield  {author} {\bibinfo {author} {\bibfnamefont {C.~W.}\ \bibnamefont
  {Galdino}}, \bibinfo {author} {\bibfnamefont {D.~C.}\ \bibnamefont
  {Freitas}}, \bibinfo {author} {\bibfnamefont {C.~P.~C.}\ \bibnamefont
  {Medrano}}, \bibinfo {author} {\bibfnamefont {D.~R.}\ \bibnamefont
  {Sanchez}}, \bibinfo {author} {\bibfnamefont {R.}~\bibnamefont {Tartaglia}},
  \bibinfo {author} {\bibfnamefont {L.~P.}\ \bibnamefont {Rabello}}, \bibinfo
  {author} {\bibfnamefont {A.~A.}\ \bibnamefont
  {Mendon\ifmmode~\mbox{\c{c}}\else \c{c}\fi{}a}}, \bibinfo {author}
  {\bibfnamefont {L.}~\bibnamefont {Ghivelder}}, \bibinfo {author}
  {\bibfnamefont {M.~A.}\ \bibnamefont {Continentino}}, \bibinfo {author}
  {\bibfnamefont {M.~J.~M.}\ \bibnamefont {Zapata}}, \bibinfo {author}
  {\bibfnamefont {C.~B.}\ \bibnamefont {Pinheiro}}, \bibinfo {author}
  {\bibfnamefont {G.~M.}\ \bibnamefont {Azevedo}}, \bibinfo {author}
  {\bibfnamefont {J.~A.}\ \bibnamefont {Rodr\'{\i}guez-Velamaz\'an}}, \bibinfo
  {author} {\bibfnamefont {G.}~\bibnamefont {Garbarino}}, \bibinfo {author}
  {\bibfnamefont {M.}~\bibnamefont {N\'u\~nez Regueiro}},\ and\ \bibinfo
  {author} {\bibfnamefont {E.}~\bibnamefont {Granado}},\ }\bibfield  {title}
  {\bibinfo {title} {Structural and spectroscopic investigation of the
  charge-ordered, short-range ordered, and disordered phases of the
  \chem{Co_3O_2BO_3} ludwigite},\ }\href
  {https://doi.org/10.1103/PhysRevB.104.195151} {\bibfield  {journal} {\bibinfo
   {journal} {Phys. Rev. B}\ }\textbf {\bibinfo {volume} {104}},\ \bibinfo
  {pages} {195151} (\bibinfo {year} {2021})}\BibitemShut {NoStop}%
\bibitem [{\citenamefont {Kazak}\ \emph {et~al.}(2022)\citenamefont {Kazak},
  \citenamefont {Arauzo}, \citenamefont {Bartolomé}, \citenamefont {Molokeev},
  \citenamefont {Dudnikov}, \citenamefont {Solovyov}, \citenamefont {Borus},\
  and\ \citenamefont {Ovchinnikov}}]{kazak2022}%
  \BibitemOpen
  \bibfield  {author} {\bibinfo {author} {\bibfnamefont {N.}~\bibnamefont
  {Kazak}}, \bibinfo {author} {\bibfnamefont {A.}~\bibnamefont {Arauzo}},
  \bibinfo {author} {\bibfnamefont {J.}~\bibnamefont {Bartolomé}}, \bibinfo
  {author} {\bibfnamefont {M.}~\bibnamefont {Molokeev}}, \bibinfo {author}
  {\bibfnamefont {V.}~\bibnamefont {Dudnikov}}, \bibinfo {author}
  {\bibfnamefont {L.}~\bibnamefont {Solovyov}}, \bibinfo {author}
  {\bibfnamefont {A.}~\bibnamefont {Borus}},\ and\ \bibinfo {author}
  {\bibfnamefont {S.}~\bibnamefont {Ovchinnikov}},\ }\bibfield  {title}
  {\bibinfo {title} {Anisotropic thermal expansion and electronic transitions
  in the \chem{Co_3BO_5} ludwigite},\ }\href
  {https://doi.org/10.1039/D2DT00270A} {\bibfield  {journal} {\bibinfo
  {journal} {Dalton Trans.}\ }\textbf {\bibinfo {volume} {51}},\ \bibinfo
  {pages} {6345} (\bibinfo {year} {2022})}\BibitemShut {NoStop}%
\bibitem [{\citenamefont {Toby}\ and\ \citenamefont
  {Von~Dreele}(2013)}]{Toby2013}%
  \BibitemOpen
  \bibfield  {author} {\bibinfo {author} {\bibfnamefont {B.~H.}\ \bibnamefont
  {Toby}}\ and\ \bibinfo {author} {\bibfnamefont {R.~B.}\ \bibnamefont
  {Von~Dreele}},\ }\bibfield  {title} {\bibinfo {title} {{{\it GSAS-II}: the
  genesis of a modern open-source all purpose crystallography software
  package}},\ }\href {https://doi.org/10.1107/S0021889813003531} {\bibfield
  {journal} {\bibinfo  {journal} {Journal of Applied Crystallography}\ }\textbf
  {\bibinfo {volume} {46}},\ \bibinfo {pages} {544} (\bibinfo {year}
  {2013})}\BibitemShut {NoStop}%
\bibitem [{\citenamefont {Shannon}(1976)}]{shannon1976revised}%
  \BibitemOpen
  \bibfield  {author} {\bibinfo {author} {\bibfnamefont {R.~D.}\ \bibnamefont
  {Shannon}},\ }\bibfield  {title} {\bibinfo {title} {Revised effective ionic
  radii and systematic studies of interatomic distances in halides and
  chalcogenides},\ }\href@noop {} {\bibfield  {journal} {\bibinfo  {journal}
  {Acta crystallographica section A: crystal physics, diffraction, theoretical
  and general crystallography}\ }\textbf {\bibinfo {volume} {32}},\ \bibinfo
  {pages} {751} (\bibinfo {year} {1976})}\BibitemShut {NoStop}%
\bibitem [{\citenamefont {Momma}\ and\ \citenamefont
  {Izumi}(2011)}]{momma2011vesta}%
  \BibitemOpen
  \bibfield  {author} {\bibinfo {author} {\bibfnamefont {K.}~\bibnamefont
  {Momma}}\ and\ \bibinfo {author} {\bibfnamefont {F.}~\bibnamefont {Izumi}},\
  }\bibfield  {title} {\bibinfo {title} {\chem{VESTA} 3 for three-dimensional
  visualization of crystal, volumetric and morphology data},\ }\href
  {https://doi.org/10.1107/S0021889811038970} {\bibfield  {journal} {\bibinfo
  {journal} {Journal of Applied Crystallography}\ }\textbf {\bibinfo {volume}
  {44}},\ \bibinfo {pages} {1272} (\bibinfo {year} {2011})}\BibitemShut
  {NoStop}%
\end{thebibliography}
%

\clearpage

\clearpage

\begin{table}[h!]
\centering
\caption{\label{hkl} Comparison between observed and calculated intensities for superstructure $hkl$ reflections at $T=525$ K. The space group is $Pbnm$ (number 62) and the lattice parameters at this temperature are $a=9.3325$ \AA, $b=11.9971$ \AA\, and $c=5.9766$ \AA. In the last two lines, the intensities of the 330 and 440 reflections that are also present in the regular ludwigite structure are given for comparison.}
\begin{ruledtabular}
\begin{tabular}{c c c c c}
$h$ & $k$ & $l$ & $I_{obs}$ & $ I_{cal}$ \\
\hline
2 & 2 & 1 & 0.018(1) & 0.019 \\
2 & 3 & 1 & 0.55(3) & 0.50 \\
2 & 4 & 1 & 0.20(1) & 0.01 \\
3 & 1 & 1 & 0.54(2) & 0.55 \\
3 & 2 & 1 & 1.32(7) & 1.27 \\
3 & 3 & 1 & 0.134(7) & 0.117 \\
3 & 3 & 3 & 0.94(5) & 0.94 \\
3 & 4 & 1 & 0.19(1) & 0.19 \\
3 & 5 & 1 & 3.7(2) & 3.8 \\
3 & 6 & 1 & 1.10(5) & 1.08 \\
4 & 1 & 1 & 4.4(2) & 4.7 \\
4 & 2 & 1 & 3.9(2) & 4.1 \\
4 & 3 & 1 & 0.122(6) & 0.123 \\
4 & 4 & 1 & 1.73(9) & 1.67 \\
4 & 5 & 1 & 0.091(5) & 0.090 \\
4 & 6 & 1 & 0.012(1) & 0.001 \\
5 & 2 & 1 & 0.63(3) & 0.58 \\
5 & 3 & 1 & 2.6(1) & 0.9 \\
\hline
3 & 3 & 0 & 570(30) & 548 \\
4 & 4 & 0 & 1610(80) & 1679 \\

\end{tabular}
\end{ruledtabular}
\end{table}

\begin{table}[h!]
\centering
\caption{\label{refined} Wyckoff sites and refined atomic positions for Co$_3$O$_{2}$BO$_3$ at $T=525$ K obtained from single crystal X-ray diffraction data. The Debye-Waller parameters are constrained to be the same for each atomic species and the refined values are $100 \times U_{iso}=1.116(16)$ \AA $^{2}$ for Co, 0.78(4) \AA $^{2}$ for O, and 0.70(9) \AA $^{2}$ for B. The structural refinement employed the superstructure reflection intensities listed in Table \ref{hkl} ($R_F^2 = 4.5 \%$) and additional 5040 observations (1480 independent $hkl$ reflections) of regular ludgiwite reflections used in a previous work \cite{galdino2021} ($R_F^2 = 9.2 \%$)} 
\begin{ruledtabular}
\begin{tabular}{c c c c c}
Atom & Wyckoff & $x$ & $y$ & $z$ \\
\hline
Co(1) & $4a$ &  1/2 & 1/2 & 0 \\   
Co(2) & $4c$ &  0.00072(5) & 0.499564(16) & 1/4 \\   
Co(3) & $8d$ &  0.00249(6) &  0.72231(5) & 0.00376(5) \\   
Co(4a) & $4c$ & 0.74060(7) & 0.38649(5) & 1/4 \\   
Co(4b) & $4c$ & 0.26020(7) & 0.61361(5) & 1/4 \\   
O(1) & $4c$ & 0.6500(4) & 0.5411(3) & 1/4 \\   
O(2) & $4c$ & 0.3428(3) & 0.4622(3) & 1/4 \\   
O(3) & $8d$ & 0.1174(3) & 0.5759(3) & 0.9831(3) \\   
O(4) & $4c$ & 0.8780(3) & 0.6421(2) & 1/4 \\  
O(5) & $4c$ & 0.1188(3) & 0.3659(3) & 1/4 \\   
O(6) & $8d$ &  0.6118(3) &  0.3567(2) & 0.0093(5) \\   
O(7) & $4c$ & 0.8417(4) & 0.2405(3) & 1/4 \\   
O(8) & $4c$ & 0.1576(3) & 0.7652(3) & 1/4 \\   
B(1) & $4c$ &  0.7394(6)  & 0.6410(4) & 1/4 \\   
B(2) & $4c$ & 0.2685(6) & 0.3731(5) & 1/4 \\   

\end{tabular}
\end{ruledtabular}
\end{table}

\begin{table*}
\centering
\caption{\label{table:distances} Relevant Co-O and Co-Co interatomic distances at $T=525$ K extracted from the data of Table \ref{refined}. All values are given in Angstroms.}
\begin{ruledtabular}
\begin{tabular}{c c c c}
Co(1)$-$O(1) & $2.106(3) \times 2$ & Co(1)$-$O(2) & $2.142(3) \times 2$ \\
Co(1)$-$O(6) & $2.012(3) \times 2$ & & \\
$<$Co$(1)-$O$>$ & 2.087(2) & & \\
\hline
Co(2)$-$O(3) & 1.994(3) $\times 2$ & Co(2)$-$O(3) & 2.138(3) $\times 2$ \\
Co(2)$-$O(4) & 2.058(3) $\times 1$ & Co(2)$-$O(5) & 1.946(4) $\times 1$\\
$<$Co$(2)-$O$>$ & 2.045(2) & & \\
\hline
Co(3)$-$O(3) & 2.061(4) $\times 1$ & Co(3)$-$O(4) & 2.108(3) $\times 1$\\
Co(3)$-$O(5) & 2.168(3) $\times 1$ & Co(3)$-$O(6) & 1.933(3) $\times 1$\\
Co(3)$-$O(7) & 2.148(3) $\times 1$ & Co(3)$-$O(8) & 2.128(3) $\times 1$\\
$<$Co$(3)-$O$>$ & 2.091(2) & & \\
\hline
Co($4a$)$-$O(1) & 2.038(4) $\times 1$ & Co($4a$)$-$O(3) & 1.975(3) $\times 2$\\
Co($4a$)$-$O(6) & 1.908(3) $\times 2$ & Co($4a$)$-$O(7) & 1.990(4) $\times 1$\\
$<$Co($4a$)$-$O$>$ & 1.966(2) & & \\
\hline
Co($4b$)$-$O(2) & 1.973(4) $\times 1$ & Co($4b$)$-$O(3) & 2.127(3) $\times 2$\\
Co($4b$)$-$O(6) & 1.989(3) $\times 2$ & Co($4b$)$-$O(8) & 2.056(4) $\times 1$\\
$<$Co($4b$)$-$O$>$ & 2.044(2) & & \\
\hline
Co$(4a)-$Co(1) & 3.0214(6) $\times 2$ &
Co$(4b)-$Co(1) & 3.0164(6) $\times 2$ \\
Co$(4a)-$Co(2) & 2.7809(8) $\times 1$ &
Co$(4b)-$Co(2) & 2.7814(8) $\times 1$ \\
Co$(4a)-$Co(3) & 3.1229(8) $\times 2$ &
Co$(4b)-$Co(3) & 3.1066(8) $\times 2$ \\
Co$(4a)-$Co(3) & 3.3455(8) $\times 2$ &
Co$(4b)-$Co(3) & 3.3598(8) $\times 2$ \\
& Co$(4a)-$Co($4b$) & 2.98831(1) $\times 2$ & \\
\end{tabular}
\end{ruledtabular}
\end{table*}

\end{document}